# Advancing atom probe tomography capabilities to understand bone microstructures at near-atomic scale

Tim M. Schwarz[1*], Maïtena Dumont[1,2], Victoria Garcia-Giner[3,4], Chanwon Jung[1,5] Alexandra E. Porter[3], Baptiste Gault[1,3*]

1. Max-Planck-Institute for Sustainable Materials, Max-Planck-Str. 1, Düsseldorf 40237, Germany
2. *now at* Groupe Physique des Matériaux, Université de Rouen, Saint Etienne du Rouvray, Normandie 76800, France
3. Department of Materials, Imperial College London, London, SW7 2AZ, UK
4. *now at* Rosalind Franklin Institute, Harwell Campus, Didcot, Oxfordshire, OX11 0QX
5. *now at* Department of Materials Science and Engineering, Pukyong National University, 45 Yongso-ro, Nam-gu, 48513 Busan, Republic of Korea

*Corresponding author E-mail address: tim.schwarz@mpie.de, b.gault@mpie.de

## Abstract

Bone structure is generally hierarchically organized into organic (collagen, proteins, ...), inorganic (hydroxyapatite (HAP)) components. However, many fundamental mechanisms of the biomineralization processes such as HAP formation, the influence of trace elements, the mineral-collagen arrangement, etc., are not clearly understood. This is partly due to the analytical challenge of simultaneously characterizing the three-dimensional (3D) structure and chemical composition of biominerals in general at the nanometer scale, which can, in principle be achieved by atom probe tomography (APT). Yet, the hierarchical structures of bone represent a critical hurdle for APT analysis in terms of sample yield and analytical resolution, particularly for trace elements, and organic components from the collagen appear to systematically get lost from the analysis. Here, we applied *in-situ* metallic coating of APT specimens within the focused ion beam (FIB) used for preparing specimens, and demonstrate that the sample yield and chemical sensitivity are tremendously improved, allowing the analysis of individual collagen fibrils and trace elements such as Mg and Na. We explored a range of measurement parameters with and without coating, in terms of analytical resolution performance and determined the best practice parameters for analyzing bone samples in APT. To decipher the complex mass spectra of the bone specimens, reference spectra from pure HAP and collagen were acquired to unambiguously identify the signals, allowing us to analyze entire collagen fibrils and interfaces at the near-atomic scale. Our results open new possibilities for understanding the hierarchical structure and chemical heterogeneity of bone structures at the near-atomic level and demonstrate the potential of this new method to provide new, unexplored insights into biomineralization processes in the future.

# 1. Introduction

The importance of the human skeleton is undisputed and includes the mechanical stabilization and protection of e.g. critical internal organs [1,2], but it also regulates several essential metabolic processes as one of our largest ion exchangers [3–5]. The skeleton is made from bone, which is hierarchically structured on different length scales from micrometers to nanometers scales and exhibits remarkable mechanical properties due to its unique combination of organic and inorganic components [1,2]. In general, there are two main constituents of bone: hydroxyapatite (HAP), which is the main mineral phase and is responsible for high stiffness, strength and wear resistance; and an organic component, responsible for the high ductility of the composite [6–9]. By weight, HAP represents 60 wt. % of the bone, 20–30 wt. % are organic components and 10 wt. % water molecules present in the collagen-mineral structure [1,10]. The majority of the organic content of bone consists of type I collagen (approx. 90 wt. %), which forms the collagen fibrils that act as building blocks for higher order structures, followed by non-collagenous proteins (e.g. proteoglycans and osteopontin) that are involved in regulating collagen development, controlling fiber size and mineralization [11].

The different processes involved in the formation of these two components are known as biomineralization, and the growth dynamics and resulting structural architecture are controversially debated [12]. In addition, rather limited information is available regarding the influence of different substitutional trace elements on the interfacial process, bonding and distribution between HAP and collagen on the nanometer scale [13], and how different substitutes influence the biomineralization and growth [14]. However, understanding the hierarchical structures of bone and the biomineralization processes at different length scales is essential for a better understanding of bone fractures and remodeling, diseases (osteolysis, osteo-imperfecta) and the assessment of pathologies. Over the last decades, intensive research has been undertaken to decipher the hierarchical structure of bone at different length scales using a wide range of microscopic techniques, including scanning electron microscopy (SEM) and (scanning) transmission electron microscopy ((S)TEM); spectroscopies including energy dispersive x-ray (EDX), electron-energy loss spectroscopy (EELS), and tomography's *via* SEM combined with a focused-ion beam (FIB), X-ray micro/nano-tomography (micro-CT) and (nano-CT); and only to name few [15–18].

Despite significant progress in elucidating the organization and structure of mineralized collagen fibrils, HAP nanocrystals and the influence of substitutional trace elements [19–23], many fundamental questions, such as where trace elements are located in the tissue (e.g. growth plate), remain unanswered to provide a better understanding of the underlying mineralization processes. This is partly due to the analytical challenge of simultaneously characterizing the three-dimensional (3D) structure and chemical composition of biominerals in general at the nanometer scale, as well as the electron dose sensitivity of bone samples.

For spectroscopic techniques such as EDX and EELS, the low atomic number and low concentration of trace elements such as Na, Mg, K, make their quantification extremely difficult. Therefore, the current shortcomings of the combined spatial and chemical resolution capabilities of analytical, spectroscopic and tomographic techniques used so far leave knowledge gaps to fully understand the biomineralization processes.

Atom probe tomography (APT) allows for 3D chemical composition mapping on the nanoscale, with an equivalent chemical sensitivity over the entire periodic table, down to the range of parts per million (ppm) [24,25]. APT is based on the field evaporation of surface atoms from a needle-shaped specimen with a radius of less than <100 nm caused by a strong electric field at the tip. Surface atoms are ionized and repelled from the charged surface and accelerated towards a position-sensitive detector [26]. To trigger the field evaporation, either high voltage or laser pulses are used in addition to an applied direct current (DC) voltage. The former lead to a fast increase in the electric field, the latter to an increase of the tip's temperature, both assisting with triggering field evaporation of the surface atoms [24,25,27]. The spatial coordinates of each ion impact are collected, and by using a time-off-flight mass spectrometry, the chemical identity of each ion can be determined. As the field evaporation proceeds, almost atom-by-atom, and layer by layer, the sequence of detection, along with the ion impact coordinates can be used, assuming a projection model, to build a three-dimensional reconstruction of the measured volume [24,28].

While APT is well established for the analysis of metallic materials [29,30] and progressively expanded to semiconductors [31], in recent years it has been extended to the analysis of different systems such as teeth [32–35], bones [13,36–38], marine organisms [39–41] and synthetic biomaterials [42–44]. However, there are still many challenges to overcome in the analysis of biominerals with APT. The heterogeneous structure of the inorganic and organic domains complicates the sample preparation, decreases the sample yield, reduces the local and chemical resolution and leads to a complex and, in most cases, ambiguous elemental or ion identification [45]. Therefore, most studies in the literature have focus on dental biominerals, especially on enamel [32–35], due to its lower organic content compared to dentine. In correlation with other techniques, APT could address some of the abovementioned questions, bringing sub-nanometer insights into the local chemistry of bones. However, these obstacles need to be overcome to establish APT in the field of biominerals. Metallic coating of APT samples has been shown in the past to improve sample yield, mass resolution and data quality [46–49], but this technique has not been applied to biomaterials with a high organic fraction, such as bones.

In this work, we first sought to optimize APT performance of bone specimens through systematic variation of the analysis parameters, and by using cryogenic specimen preparation to limit beam damage to maintain the structural integrity of the sample. These do not lead to the necessary improvement in mass resolution. We, introduce an *in-situ* metallic coating of APT bone specimens, readily inside the FIB. The metallic coating significantly improves sample yield and chemical resolution performance, allowing the analysis of collagen-related signals and trace elements such as Mg and Na. We varied measurement parameters with and without coating and report the analytical performance, and determined the best practice parameters for the analysis of bone samples in APT. We highlight the tremendous effect of the metallic coating on the analytical performance. To help decipher the complex mass spectra of the bone samples, spectra were acquired from reference samples, pure HAP and collagen, to unambiguously identify sets of peaks, allowing us to analyze the complete collagen fibrils and interfaces. Our results open new possibilities for understanding the hierarchical assembly and chemical heterogeneity of bone structures at the near-atomic level and demonstrate the potential of this new *in-situ* coating method to facilitate and improve APT analysis of biominerals and, in the future, provide new insights into biomineralization processes, collagen chemistry/structure and changes in pathologic tissues.

# 2. Materials and Methods

## 2.1. Materials

Pure Cr metal (99.99%) from Goodfellow was used for *in-situ* sputtering/coating of the APT specimens. The metal flakes were cut into smaller pieces where necessary, then ground and polished (up to 1 μm) to create a flat surface and to remove the oxidized surface. Flakes were mounted onto a SEM stub using adhesive copper tape.

As a reference for the inorganic HAP phase of the bone measurement, APT specimens of pure HAP powder with the particle sizes of 10 μm, purchased from Sigma Aldrich, were prepared and used to compare mass spectra and find any fingerprint signals of HAP (CAS-number 1306-06-5). The powder was dispersed on a carbon tape previously mounted on a SEM stub, and the excess powder was removed with pressurized nitrogen gas. The same protocol was used to make APT specimens of pure bovine collagen type I powder, purchased from Sigma Aldrich (CAS-number 9007-34-5) and used as a reference for the organic phase of the bone. The sample was then stored in a refrigerator at 5 °C until the APT specimens were prepared.

The bone tissue samples from an adult female mouse (+ 6months, C57BL/6J/Bcl2 OE) used in this research were donated by the Imperial College CBS facility and sourced from animals being hand-on for another non-skeletal research.

## 2.2. Bone sample preparation

The femurs were dissected by careful removal of surrounding soft tissue to isolate the bone tissue. Sample were immediately stored in phosphate-buffered saline (PBS) buffer-soaked gauze and stored at -20°C until further isolation of mid-shaft femora.

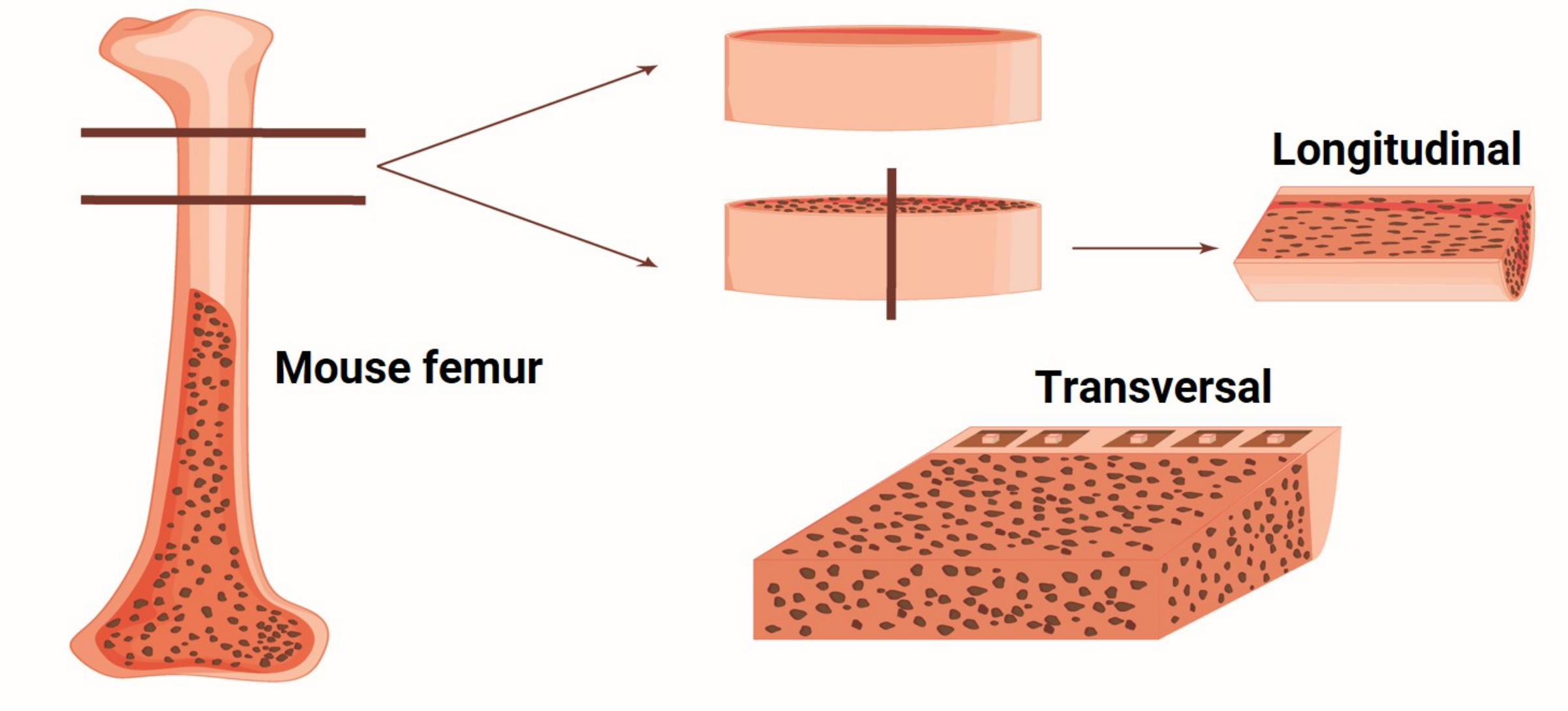

**Figure 1**: Schematic sketch of the sectioning and sample preparation of the mouse femur.

Cortical bone tissue specimen preparation was adapted from a previous protocol by Aldegaither et al. [50]. Femurs were thawed and sectioned to approximately 500 µm, using a Buehler IsoMet Low Speed Saw equipped with a diamond-coated blade (15HC 0,15mm) at 60-100 rpm. A counterweight was applied to ensure consistent pressure during cutting, while PBS was continuously used to cool the blade and keep the bone sample hydrated, minimizing heat generation and preventing sample dehydration. To facilitate cutting, a small portion of the bone was mounted in a fast, cold-curing resin (Buehler Varidur 10), leaving the femur shaft and head exposed to allow for precise sectioning. The schematic in **Figure 1** shows the sectioning of the cortical bone sample.

Sectioned femurs with a diameter of 1-2 mm were mounted on SEM stub using carbon tape and silver paste. Samples were wet ground with super fine SiC paper (P4000), polished with diamond paste (3 to 0.25 µm) and washed in buffer to obtain a flat surface. The samples were carbon-coated (Leica ACE600) to minimize material charging during exposure to the electron beam. Samples were stored at room temperature until FIB milling.

## 2.3. Atom probe specimen preparation

For the APT specimen preparation a Thermo-Fisher 5 CX Ga SEM-FIB equipped with an Aquilos cryo-stage with free rotation capability was used [49]. The needle-shaped

APT specimens were prepared using an FIB lift-out protocol based on the sample preparation protocol of Thompson et al. [51]. A detailed step-by-step sample preparation of the bone sample is shown in **Figure S1**. The lamella was prepared away from any blood vessels and osteocytes. To protect the sample surface during preparation and to avoid any structural changes/damage to the internal material, a Pt protective layer (20 x 2.5 x 1 µm) was first deposited on the surface using the gas injection system (GIS), followed by cutting three trenches into the material. Using a micromanipulator, the lamella was lifted out of the material and then attached to Si support posts (Cameca Scientific Instruments). To compare APT specimens prepared at room temperature (RT) and cryogenic temperatures, the FIB stage was cooled to -190°C by circulating a 190 mg/s $N_2$ gas flow through a heat exchanger system inside a liquid nitrogen dewar after the lamellae were mounted on the Si support post. In both cases the samples were tilted to 52°, followed by annular milling with decreasing inner radius and current until a tip diameter of r < 100 nm was achieved. Finally, a pattern at 5 kV, 7 pA was used for several seconds to remove regions implanted with Ga or possible amorphized, damaged surface layer. For electron imaging of the specimens, especially at higher magnification, low currents and acceleration voltages were used to avoid bending of the final APT specimen.

Cr was chosen for the *in-situ* metallic coating of the APT specimens of the bone specimens. A lift-out of pure Cr was prepared according to the protocol described above, followed by cutting a semicircle in the lamella. For *in-situ* coating, the x, y direction of the lamella was positioned in the electron beam and the z direction in the ion beam to be above the tip of the specimen. The specimens were then coated with Cr for 4 x 25s at 40 pA. A detailed description of the *in-situ* coating process can be found in Schwarz et al. [48].

The specimens were then transferred at RT for further analysis, either to a LEAP 5000 XR (reflectron, Cameca Scientific Instruments) atom probe for higher mass resolution or to a LEAP 5000 XS (straight path).

## 2.4. Atom probe measurement, data reconstruction and processing

The APT experiments were performed with the LEAP 5000 XR atom probe for the parameter study and a LEAP 5000 XS to analyze the dissociation of evaporated molecular ions by using correlation plots [52]. The APT measurements were performed in laser pulsing mode at a base temperature of 40 K. Details of the laser pulse energy, laser frequency and detection rates will be given below.

Data reconstruction was performed using Cameca's Integrated Visualization and Analysis Software (IVAS) in AP Suite 6.3. For all reconstructions, the specimen radius evolution model was used, which is based on the assumption of a constant shank angle [53]. The initial radius and shaft angle were determined from the SEM images. A field factor of $k = 3.3$ and an image compression factor of $\xi = 1.65$ were used in all reconstructions. All identified signals for the uncoated and Cr coated specimens are listed in **Table S1**.

## 2.5. Scanning transmission electron microscopy

TEM sample preparation was performed under RT using the protocol in Ref. [54], whereby thinning of the lamella was performed at cryogenic temperature to limit or avoid structural damage during TEM sample preparation. GIS-based Pt was first deposited on the targeted area with electron-beam and afterwards with the ion-beam to protect surface at RT. The lamella was extracted using the micromanipulator and mounted on a TEM copper grid. To thin the lamella, the FIB was cooled down to cryogenic temperatures and the stage was rotated to 52° and the sample was tilted +/-2° in both directions, decreasing the thickness and current until the lamella reached a thickness below 100 nm. To minimize the amorphous area, the stage was tilted +/-3.5° and a rectangle at 5 kV, 7 pA was used to remove any remaining amorphous damaged areas. Finally, the lamella was heated up to RT in the FIB before being mounted in the TEM holder for further analysis.

Transmission electron microscopy (TEM) images were obtained using a Titan Themis (60–300 kV) microscope (Thermo Fisher Scientific) equipped with an image corrector, operated at 300 kV. Images and selected area electron diffraction (SAED) patterns were obtained on a CMOS (metal-oxide-semiconductor) camera (4k × 4k pixels). Scanning transmission electron micrographs were acquired on the same instrument, also operated at 300 kV. For high-angle annular dark field (HAADF) imaging, a camera length of 160 mm was used, corresponding to a collection angle of 49–200 mrad.

## 2.6. Statistical analysis

For the statistical analysis used in this study, the mean and standard deviation (±) were used e.g., for determining collagen fibril diameter from the TEM data, calculating the composition for all elements listed in **Table 1**, and calculating the amount of trace elements in **Table 3**, unless otherwise indicated. The error in the 1D composition profiles was calculated using the standard deviation of the concentration of each element/molecule type in each bin.

# 3. Results

## 3.1. Parameter sweep for uncoated specimens

Due to the heterogenous structure of bone, the influence of different measurement parameters on the measured Ca/P, overall composition, mass resolution, background, and the effective electric field derived from the charge state ratio (CSR), were analyzed for uncoated bone specimens (**Figure S2**-**Figure S5**). For this purpose, laser pulse energy, laser pulse frequency and detection rate (DR) were varied on the same specimen to limit effects due to different specimen geometries such as shank angle, radius and length. The same peak identification (see **Table S1**) and set of mass ranges were used for all analyses.

We observed loose correlations between the variation of the different measurement parameters, **Figure S2**, which can be attributed to the inhomogeneous structure of bone and the presence of two different phases with different evaporation fields. The Ca/P ratio scales with the intensity of the electric field reflected by the higher CSR of $PO_3^{2+}/PO_3^{+}$. Increasing the laser pulse energy also leads to a higher Ca/P ratio, which is more unexpected as this should result in a higher sample temperature at the tip of the specimen and hence a relatively lower electrostatic field required to maintain a constant detection rate. There is a correlation between the laser pulse energy and the organic to inorganic ratio. A high electric field is required to increase this ratio, which are usually underrepresented in APT measurements of biominerals, and especially for bone. However, a higher electric field has a negative effect on sample yield, as it causes Maxwell stresses that typically leads to earlier failure [55,56]. Lastly, a correlation has been found between laser energies and the organic to inorganic fraction, which increases linearly with higher laser energies. A mass spectrum with optimized measurement parameters is shown in **Figure 2**, with the corresponding calcium, phosphorus, oxygen and organic signals highlighted in color. A total of 96 peaks were identified and are listed in **Table S1**.

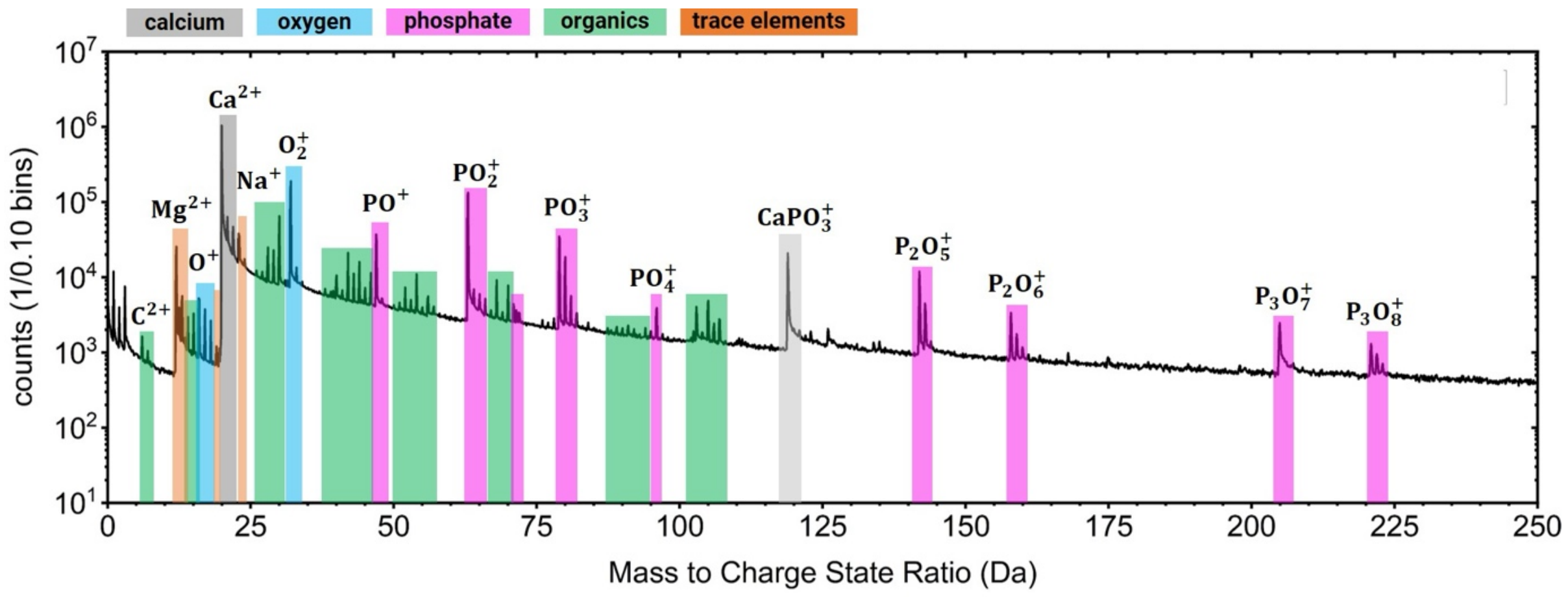

**Figure 2**: The mass spectrum of the uncoated mouse femur specimen with optimized parameters (10pJ @ 50kHz) is plotted on a logarithmic scale. The corresponding signals of calcium, phosphate, oxygen and organic related signals are highlighted in color, as well as different trace elements observed. Significant peaks are labelled, and all identified signals are listed in **Table S1**.

Based on this parameter sweep, optimization of the measurement parameters was attempted with respect to several factors such as mass resolution (FWHM), background level, Ca/P ratio, O/(Ca+P) ratio, CSR and organic fraction. The laser pulse frequency of 50 kHz was chosen to increase the time between the laser pulses and thus minimize the thermal tailing of the $Ca^{2+}$ signal, and the laser energy of 10 pJ was chosen to increase the mass resolution and reduce the thermal increase of the base temperature due to the laser pulses. A detection rate of 0.005 ions per pulse was used. The optimized parameters were intended to improve the ability to identify organic signals or trace elements below the thermal tail of the $^{40}Ca^{2+}$ signal and the general background. However, even with these optimized parameters, the tailing and background could not be completely suppressed, which led to the loss of important information on trace elements and organic signals.

## 3.2. Effect of cryo-preparation

In all previous studies, specimens from biominerals were prepared using a FIB lift-out process at room temperature [13,36,38]. However, the preparation of composite materials between a hard HAP and a soft collagen phase can cause difficulties due to the different milling behavior. For the preparation of soft materials in FIB, the energetic electron and ion beams can cause different types of radiation damage. Beam-induced heating increases the temperature of the sample due to phonon collision and depends on the ion acceleration voltage, current and thermal conductivity of the sample, which can lead to melting of the sample during preparation [57]. Bone has a poor thermal conductivity of 0.26–0.53 W/m·K [58–60], and high currents during sample preparation have been observed to mill the sample even in the inner circle pattern in the absence of the Ga-beam. At extremely high magnification in the electron beam, the sample begins to bend due to the damage caused by the radiation in the material, which reduces the mechanical strength. The same observation was made by Eder et al. [41]. In addition to thermal damage, the electron and ion beams can also cause radiolysis damage to the collagen structure, leading to a change in the chemical structure such as functional groups, chemical coordination or oxidation state, cleavage of chemical bonds or cross-linking reactions of side chains. The effect of sample preparation in FIB on the preservation of the original chemical structure has been poorly studied [57]. The use of a higher beam acceleration voltage, up to 25 kV, has been shown to reduce the amount of Ga implanted, thereby preserving the original structure [61]. This observation is controversial to the generally accepted understanding that low ion beam currents reduce Ga implantation [62,63].

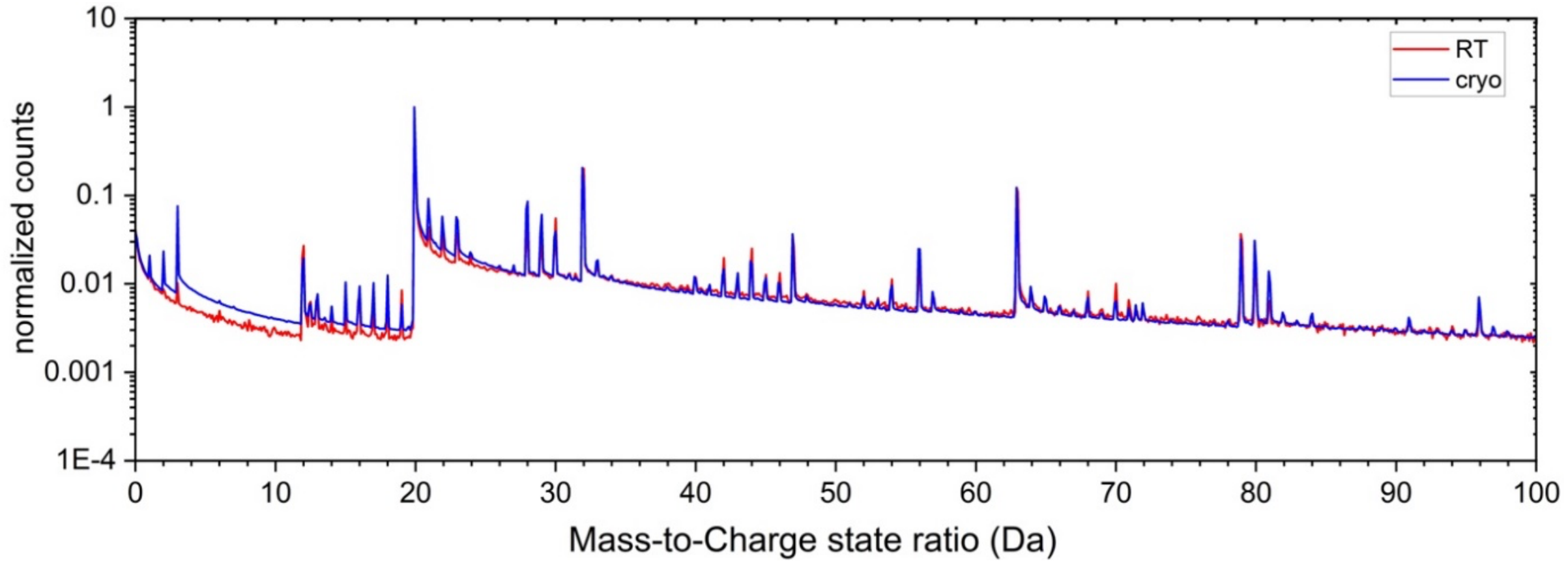

**Figure 3**: Comparison of the mass spectra prepared under RT and cryogenic temperatures. Both spectra are normalized to the highest peak.

Cryogenic sample preparation is relatively new for APT [64], whereas it is better established for TEM lamellae [65,66]. Preparation at cryo temperature is expected to help preserve the pristine state of the sample, allowing the analysis of e.g. radiation-sensitive battery materials [67] along with biological and frozen samples [68–70]. Here, specimens from the same bone were prepared at room and cryogenic temperatures. Above 510K, collagen undergoes thermal degradation and the initial structural changes [71,72]. Both spectra are normalized to the highest signal. The two mass spectra are identical (**Figure 3**), the number of detected signals, Ca/P and the composition and organic content show no significant differences. It can be concluded that cryo-preparation has no effect on sample yield, mass resolution and composition for bone samples. Consequently, it is not necessary to prepare bone samples for APT analysis at cryogenic temperatures, thus simplifying sample preparation.

### 3.3. Influence of Cr coating

The protocol for coating APT specimens was introduced in Refs. [48,49]. Herein, we provide a summary specifically for coating biomineral (bone) APT specimens *in-situ* (**Figure 4**). The thin metallic layer (10–20 nm) helps reduce the background and thermal tailing that appear, in general, due to poor electrical and thermal conductivity. A semi-circle was cut in a previously lifted-out Cr-lamella, **Figure 4A**.

This Cr-target is then positioned over the specimen by controlling both the electron- (x-, y-direction) and ion-image (z-direction) in the SEM-FIB, **Figure 4A, B**. Prior the coating, a strong charging effect can be observed in the electron image, **Figure 4C**, which is typical of poorly conducting specimens in the SEM. After coating, **Figure 4D** charging is no longer observed, a good indication of an enhanced electric charge transport through capping with the conductive Cr layer. Following reconstruction of the APT data, **Figure 4E**, atoms of Cr (in red) can be seen enveloping the APT specimen, demonstrating successfully metallic coating of the bone specimen with a sufficient adhesion.

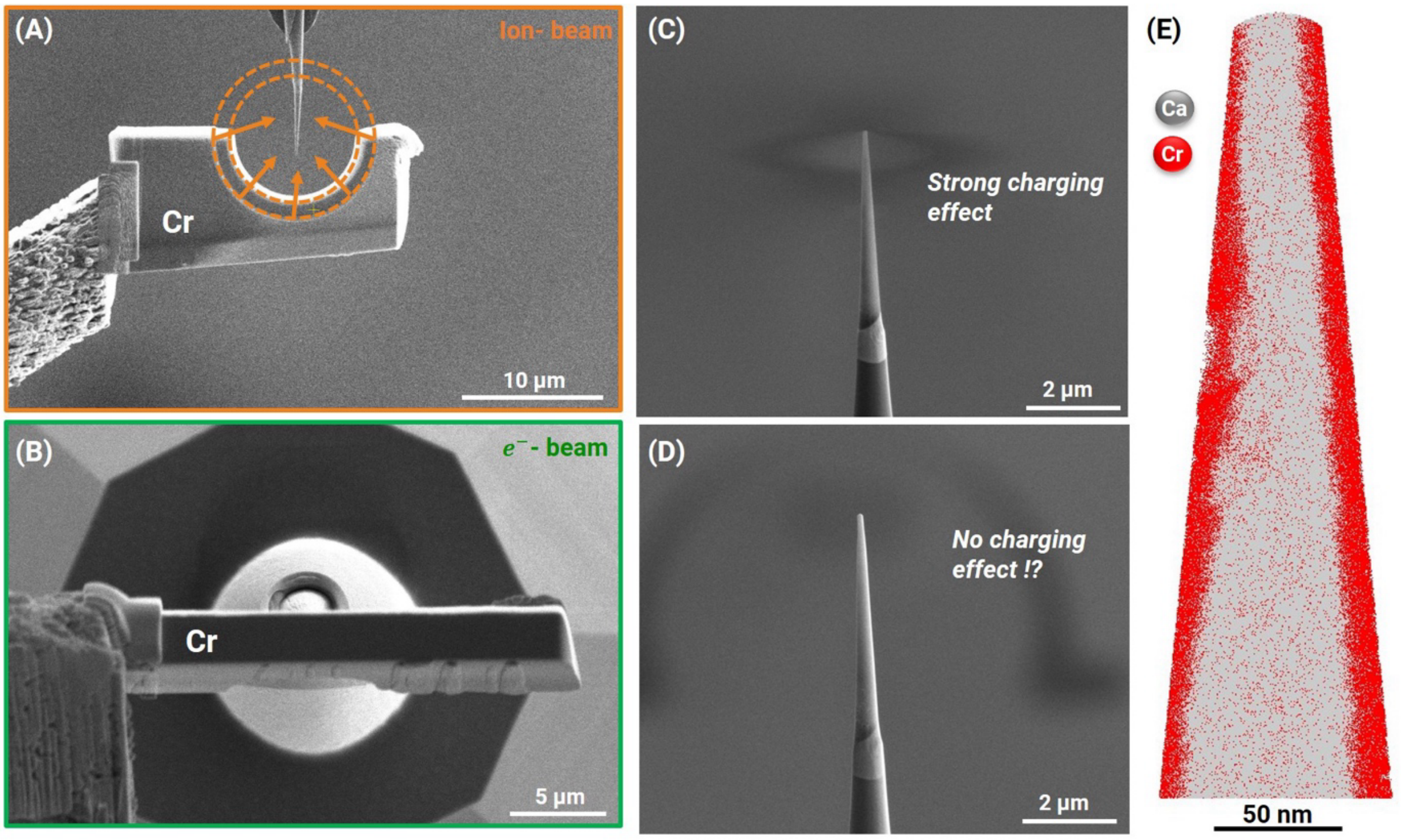

**Figure 4**: (**A, B**) *In-situ* approach to coating bone specimens with a metallic layer. (**C, D**) Comparison of the uncoated and coated APT samples shows that the charging effect due to poor electrical conductivity in the electron image disappears due to the metallic coating. (**E**) 3D reconstruction of the coated APT sample shows complete coverage of Cr (red atoms) around the entire tip.

To compare mass spectra for an uncoated and a Cr-coated specimen the same measurement parameter (40 K, 20 pJ, 125k Hz, 0.5% DR) were used, mass spectra were normalized to the highest peak, $^{40}Ca^{2+}$ at 20 Da, **Figure 5**. It is reported that the *in-situ* coating of Cr introduces Ga and oxygen into the sputtered metallic layer. Therefore, the mass spectrum of the Cr coated layer may contain signals from $CrO_x$ and $Cr_2O_x$. Further details can be found in Schwarz et al. [48]. For the comparison of the two mass spectra, the area of the bone specimen was cropped to minimize signals originating from the Cr coating, which would complicate the unambiguous interpretation of the signals. Both samples have the same specimen length and comparable tip shank angle and radius to avoid any influence from the specimen geometry. All identified signals are given in **Table S1**. The first observation is that the background in the Cr-coated specimen is reduced by 1–2 orders of magnitudes, allowing to detect more signals than in the uncoated specimen. Secondly, the thermal tailing of the $^{40}Ca^{2+}$ signal at 20 Da is significantly reduced for the Cr-coated specimen. The background level before the $^{40}Ca^{2+}$ signal at 20 Da is reached again after 4.5 Da for the Cr-coated specimen, whereas the background level for the uncoated specimen returns after 68.5 Da. These two observations indicate that the metallic coating provides better thermal conductivity, which significantly reduces the background and thermal tailing, allowing the detection of low levels of trace elements. The intensity of the electric field, as reflected by the higher CSR of $PO_3^{2+}/PO_3^{+}$, also decreased from 40.09 ± 0.25 % for the uncoated specimen to 7.52 ± 1.10 % for the Cr-coated specimen, indicating a better electric conductivity through the metallic coating. 69 signal could be identified for the uncoated specimen, while up to 120 were identified for the Cr-coated specimen. The increased number of signals can be attributed to organic signals, which have so far been underrepresented in reports of biomineral analysis by APT, including in our uncoated specimen. The metallic coating improves tremendously the detectability of trace elements and organic components.

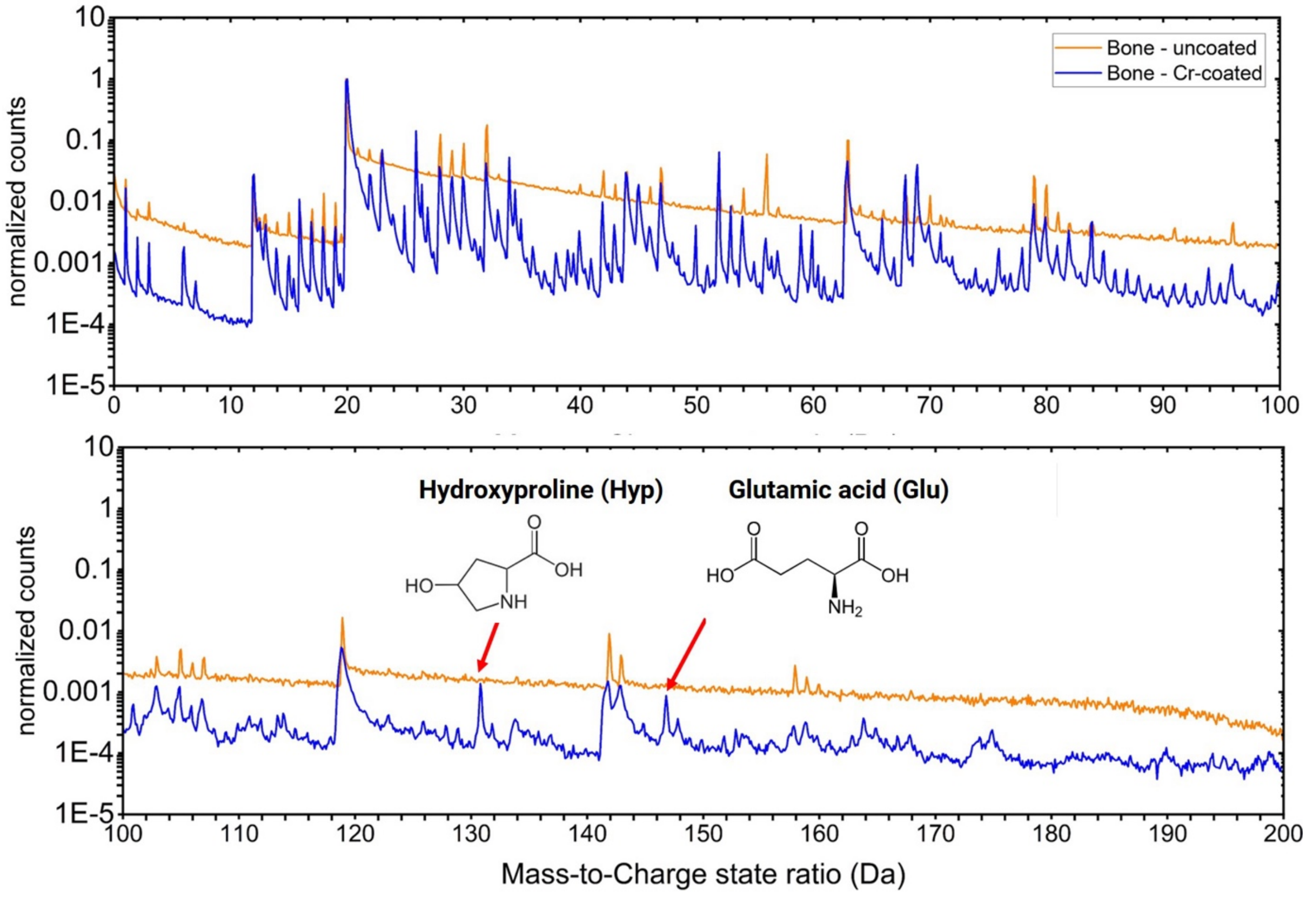

**Figure 5**: Comparison of the mass spectra of an uncoated and a Cr coated sample in logarithmic scale (note: for the comparison of the mass spectra, the area of the bone sample has been cropped to minimize signals originating from the Cr coating). Both mass spectra are normalized to the $^{40}Ca^{2+}$ signal. The presence of fully intact amino acids of hydroxyproline and glutamic acid in the mass spectra of the Cr-coated specimen are highlighted at 131 and respectively 147 Da.

Further analysis of the composition of the uncoated and Cr-coated specimens shows that the Ca/P ratio is 1.85 ± 0.11 for the uncoated specimen and decreases to 1.54 ± 0.04 for the Cr-coated specimen, which is closer to the theoretical Ca/P ratio of 1.67 for HAP (**Table 1**). The O/P ratio shows a larger deviation from the theoretical value of 4.33 than the Ca/P ratio for both specimens uncoated and coated. The reason for this larger deviation is that common amino acids (glycine (Gly), hydroxyproline (Hyp)) in the collagen have between two and four oxygen molecules per molecule and therefore lead to an overestimated oxygen concentration to measure the theoretical ratio of O/P for HAP.

**Table 1**: Composition analysis between uncoated and Cr coated bone sample. The errors were calculated with the standard deviation.

| | **Ca (at.%)** | **O (at.%)** | **P (at.%)** | **Na (at.%)** | **Mg (at.%)** | **C (at.%)** | **N (at.%)** | **H (at.%)** | **Ca/P ratio** | **O/P ratio** | **Inorganic fraction (%)** | **Organic fraction (%)** |
|---|---|---|---|---|---|---|---|---|---|---|---|---|
| **theoretical** | | | | | | | | | 1.67 | 4.33 | 70.00 | 30.00 |
| **Uncoated** | 20.63± 1.93 | 39.65± 0.26 | 11.14± 0.37 | 1.27± 0.04 | 1.07± 0.04 | 9.99± 0.72 | 3.81± 0.59 | 12.43± 1.26 | 1.85± 0.11 | 3.56± 0.10 | 85.93± 1.12 | 14.07± 1.12 |
| **Coated** | 15.38± 0.14 | 37.81± 1.11 | 10.11± 0.34 | 1.63± 0.85 | 1.26± 0.44 | 14.77± 0.47 | 3.40± 1.01 | 15.73± 0.84 | 1.54± 0.04 | 3.78± 0.02 | 78.86± 1.99 | 21.14± 1.99 |

The possibility of detecting more signals from organic components in the Cr-coated specimen affects the ratio of organic to inorganic components, with the ratio increasing to 21.14±1.99% for the Cr-coated specimen compared to 14.07±1.12% for the uncoated specimen (**Table 1**). It can therefore be concluded that a large proportion of the organic signals are lost due to the thermal tailing and higher background of the uncoated specimen. Interestingly, at higher masses, 131 Da and 147 Da, signals of hydroxyproline and glutamic acid are observed in their ionic form in the Cr-coated specimen (**Figure 5**). Both are common amino acids in collagen [73].

**Table 2**: Isotope ratio analysis of the $Ca^{2+}$ signal for the uncoated and Cr-coated specimen. The asterisk indicates the overlap of the $^{23}Na^{+}$ peak, resulting in a deviation of the isotopic ratio from the natural abundance.

| | **20 Da** | **21 Da** | **21.5 Da** | **22 Da** | **23 Da *** | **24 Da** |
|---|---|---|---|---|---|---|
| **theoretical abundance (%)** | 96.941 | 0.647 | 0.135 | 2.086 | 0.004 | 0.187 |
| **Uncoated (%)** | 73.824 | 7.676 | 4.092 | 6.613 | 6.003 | 1.792 |
| **Coated (%)** | 92.306 | 0.939 | 0.234 | 2.318 | 3.974 | 0.234 |

The Cr-coating significantly improves the mass resolution, enabling the detection of individual isotopes of $Ca^{2+}$ beyond $^{40}Ca$, i.e. $^{42}Ca$, $^{43}Ca$, $^{44}Ca$ and $^{48}Ca$, with their natural abundances reported in **Table 2**, as previously noted [48]. For the uncoated specimen, the relative $^{40}Ca^{2+}$ peak is only approx. 73.8 at. %, whereas for the coated specimen it is 92.3 at. %, noting that the natural abundance is 96.9 at. %, because the range defined for each isotope contains primarily ions from the tail of the $^{40}Ca^{2+}$ peak. It should be noted that the peak at 23 Da, $^{46}Ca^{2+}$ is higher in both measurements, due to an overlap with the $^{23}Na^{+}$ signal. The ability to detect and quantify low abundance isotopes is crucial for the unambiguous detection of trace elements in biominerals.

## 3.4. Comparison between Bone/Collagen/HAP

To support the interpretation of the additional signals observed in the Cr-coated bone specimen compared to the uncoated specimen, **Figure 5**, we performed additional measurements of pure HAP and collagen. The HAP and collagen specimens are not metallic coated with Cr. Mass spectra of collagen, HAP (both uncoated), uncoated bone and Cr-coated bone are plotted for comparison in **Figure 6**. All spectra have been normalized to the highest peak. Note that the y-axis has been adjusted to facilitate visualization. All identified peaks in the mass spectra of the pure HAP, in black in **Figure 6**, are listed in **Table S2**, corresponding to calcium, phosphate and oxygen ions, which can also be identified in the mass spectrum of the bone.

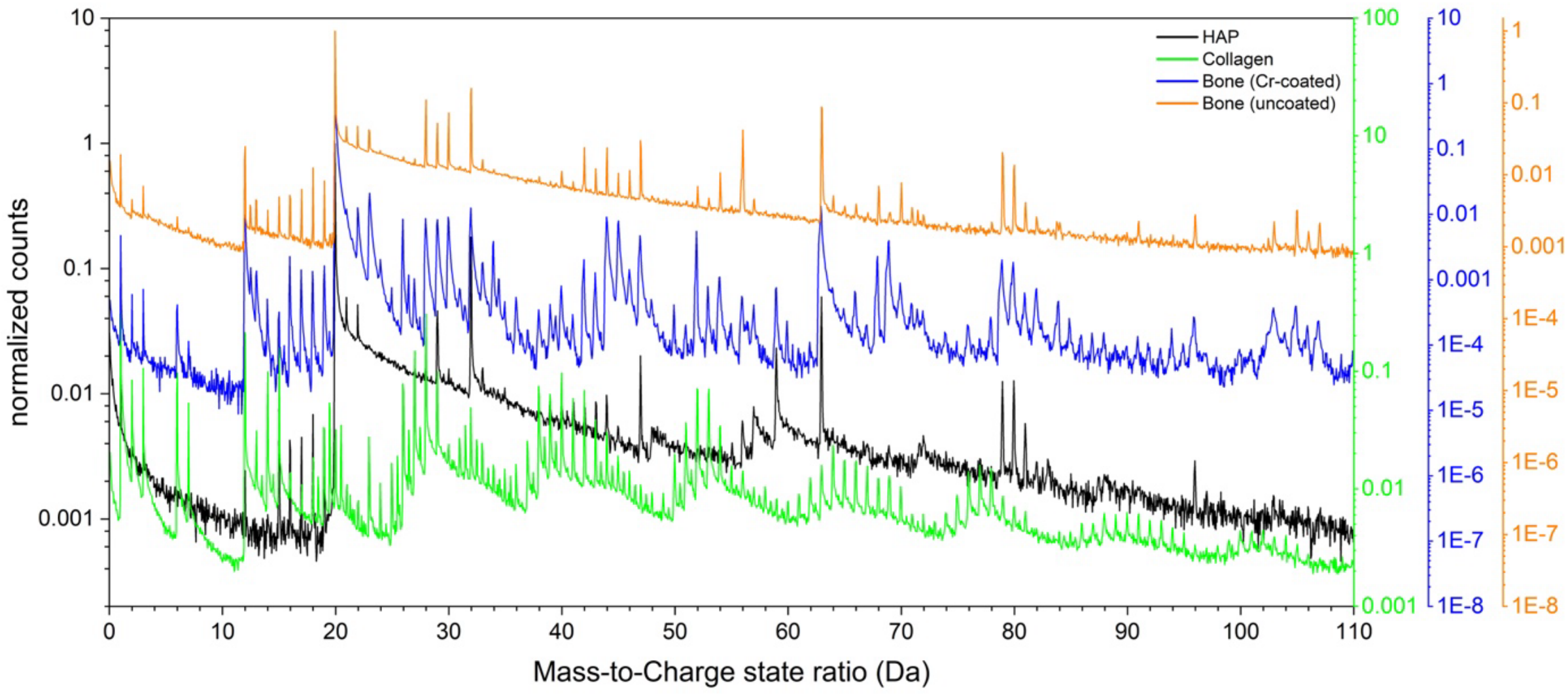

**Figure 6**: Comparison of pure collagen and HAP mass spectra (both uncoated) with uncoated and coated femur samples is plotted logarithmically. All mass spectra are normalized to the highest peak. The y-axis has been adapted to the different spectra for better visualization.

The mass spectrum of pure collagen, in green in **Figure 6**, is complex with nearly a signal at every Da above 12 Da. This fragmentation behavior is characteristic of organic molecules analyzed by APT [69,70,74–77]. The origin of many of the signals in the mass spectrum of the Cr-coated bone specimen can be attributed to collagen. In particular, the peak at 28 Da is very prominent in both the collagen and bone mass spectra and is not present in the HAP mass spectrum. Peak identification in the analysis of organics by APT is not always unambiguous, as there are potentially many combinations considering e.g. C, N, O and H atoms matching the same measured mass-to-charge state ratio. For instance, the peak at 28 Da plays an important role in the identification and localization of collagen fibrils, and can be labelled as $CO^+$, $N_2^+$, $CH_2N^+$ or $C_2H_4^+$. This comparison of mass spectra also helps clarify the localization of the collagen fibrils in the 3D reconstruction of the APT analysis of bone.

## 3.5. Structural information of collagen fibrils

STEM-HAADF images were acquired to compare the structural information with the APT reconstructions and to adjust the reconstruction parameters if necessary. The TEM lamella of the mouse femur sample was prepared at cryogenic temperatures to minimize damage to the original structure. In STEM mode, **Figure 7A-C**, bright regions indicate the presence of elements with higher atomic numbers within these areas. **Figure 7**(**A-C**) shows different orientations of fibers, the orientation of the fibers (lamellar or distorted) are depending on the area where the TEM lamellae were prepared, near the lamella structure or the Harversian canal. The gap areas between the collagen fibers have higher contrast due to the inclusion of HAP crystals. The organic rich phases (collagen) appear with dark contrast in the STEM-HAADF images due to their composition of low atomic number elements.

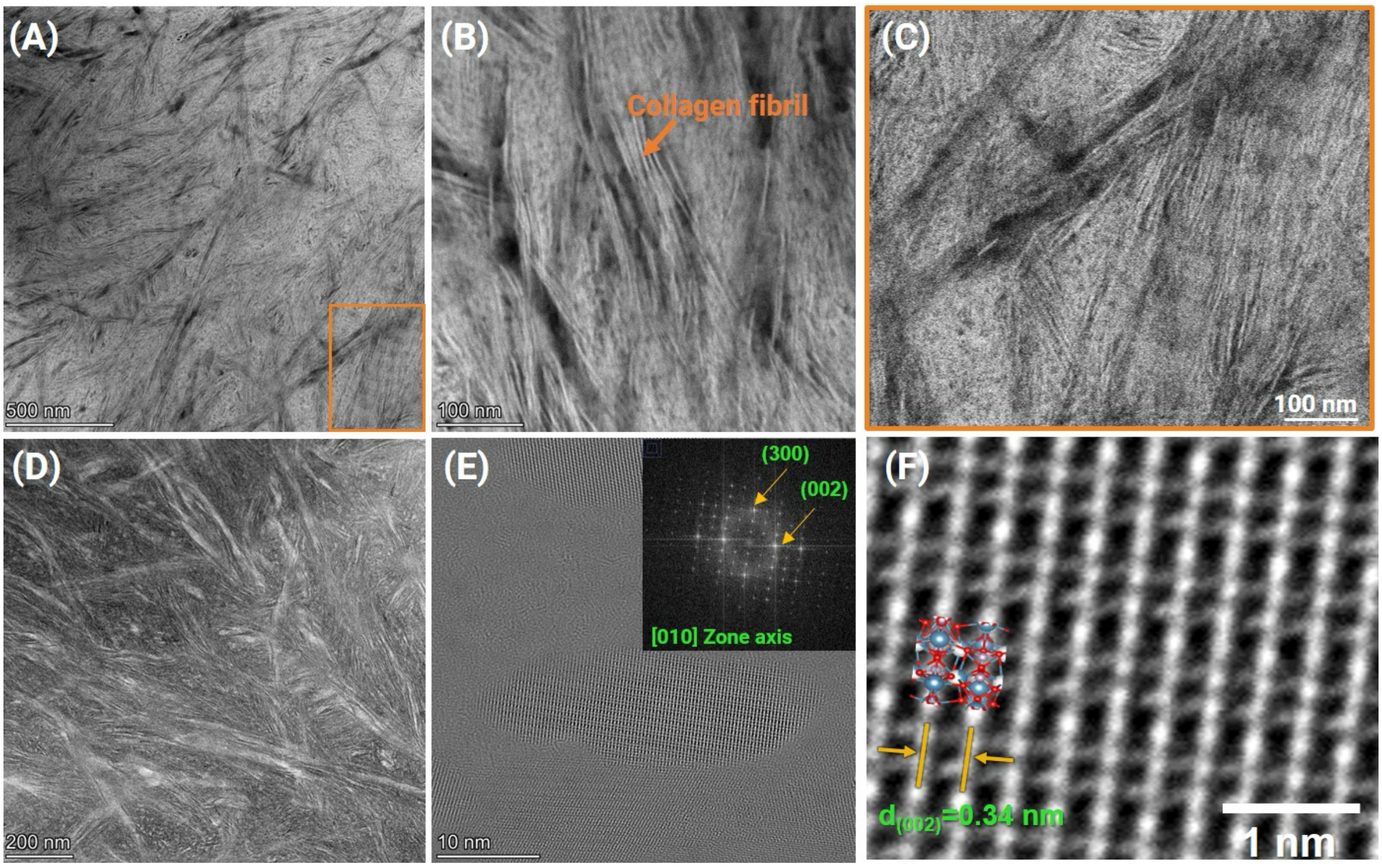

**Figure 7**: (**A-C**) STEM-HAADF images showing collagen arrangement and microfibril formation. (**D**) Bright-field TEM (BF-TEM) image and (E, F) HR-TEM images and fast Fourier transform (FFT) pattern confirm the hexagonal crystal structure of HAP.

Plate-like features are visible at higher magnification in **Figure 7B** indicating that mineralized Ca crystals aligned along the collagen fibrils. The mineralized collagen fibrils have a measured average diameter of 3.67 ± 0.67 nm (n=20), **Figure 7B**. In the BF-TEM mode (**Figure 7D**), we observe similar features with opposite contrast, resulting from differences in incident beam intensity between the two phases. Furthermore, high-resolution TEM (HR-TEM) images and fast Fourier transform (FFT) pattern reveal the hexagonal crystal structure (hcp) structure of the mineralized crystals with a P63/m symmetry and confirm the appearance of HAP crystals.

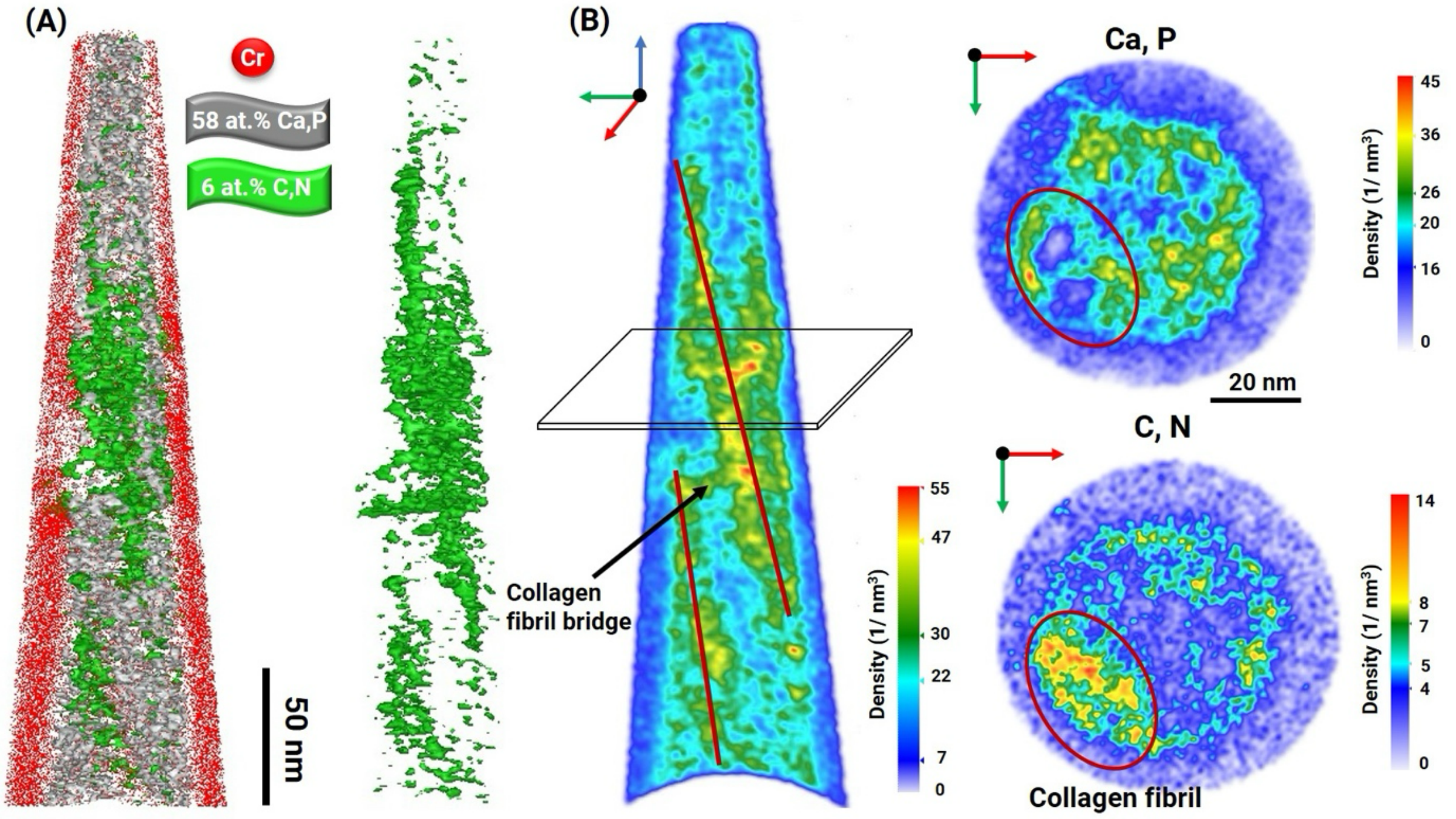

**Figure 8**: (A) 3D reconstruction of the Cr-coated bone sample, the Cr layer is homogeneously deposited around the entire sample (red atoms). The iso-concentration surfaces of Ca+P (>58 at. %) and C+N (>6 at. %) visualize the organic-rich and mineral-rich regions, respectively. (B) The density gradient map of C+N in the cross section (x, z-plane) of the sample visualizes the parallel arrangement of collagen fibrils. The density maps in the marked section (x, y-plane) show that the regions of the collagen fibrils are enriched in organic components and depleted in Ca+P.

**Figure 8** shows a 3D reconstruction from an APT analysis of a Cr-coated bone specimen. The homogeneous layer of Cr atoms (in red) around the entire specimen demonstrates uniform coating. Sets of iso-concentration surfaces in **Figure 8A**, with thresholds of Ca+P of 58 at. % in grey and C+N of 6 at. % in green, delineate the mineral-rich and the organic-rich regions respectively. The collagen microfibril arrangement can also be visualized using the point density map of C+N in the cross section (x-z-plane) across the entire dataset, **Figure 8B**. This evidences that collagen fibrils are arranged parallel to each other, which agrees with STEM as in **Figure 7B**, and bridges between the collagen fibrils could be observed, which has been reported previously [78]. The density maps in the x-y-plane, plotted in **Figure 8B**, show that the organic rich areas are deficient in Ca/P and *vice versa*.

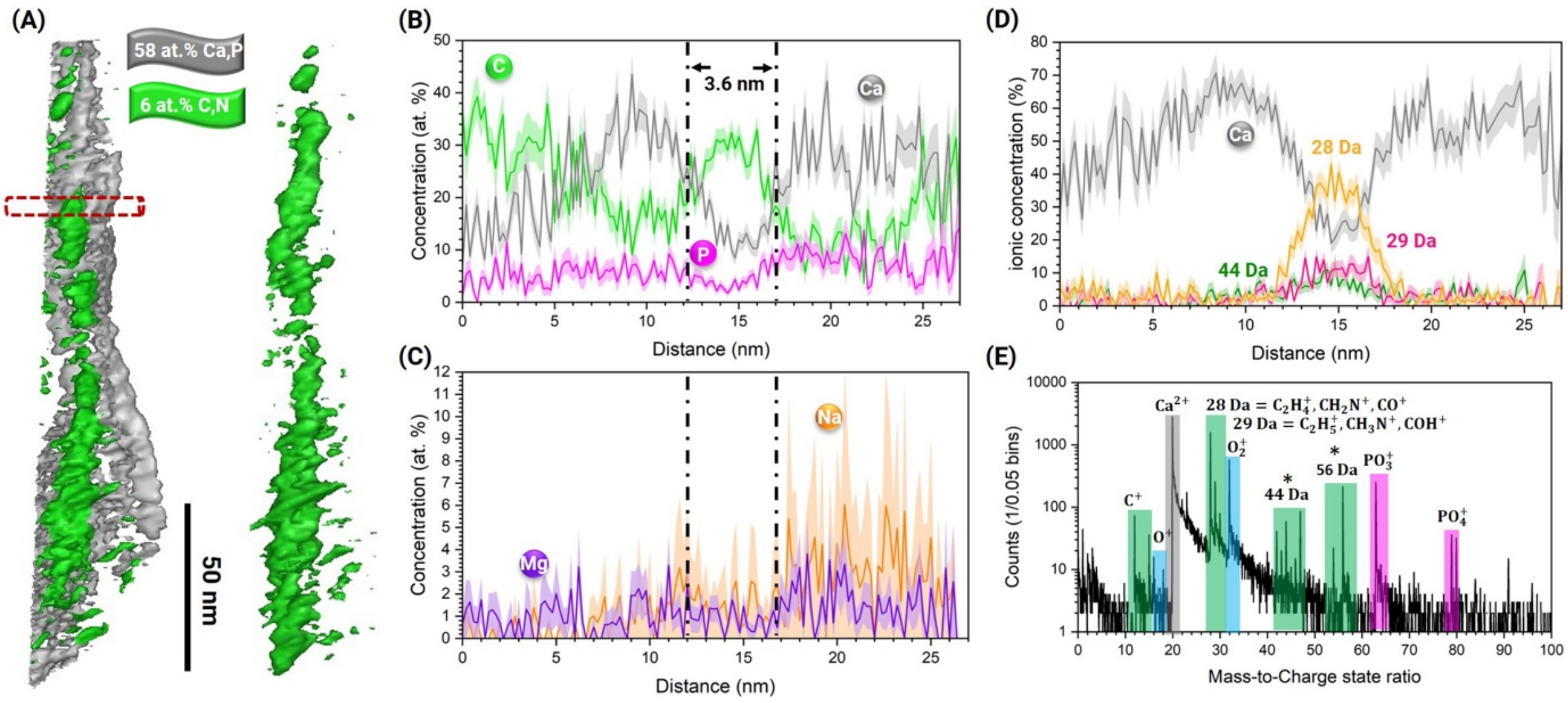

**Figure 9**: (**A**) Extracted collagen fibril from the 3D reconstruction of a second sample. The iso-concentration surface > 6 at. % of C+N visualizes the collagen fibril. (**B**) The atomic composition profile through the collagen fibril shows that the collagen fibril is enriched in the organic constituents. (**C**) Shows that the trace elements Mg and Na accumulate at the interfaces between HAP and collagen fibrils and in the HAP phase. (**D**) The ionic composition profile shows that the organic signals are derived from the signals at 28, 29 and 44 Da. (**E**) shows the mass spectrum within the fibrils, with one of the most prominent signals coming from the organic signals at 12, 28, 44 and 56 Da.

**Figure 9A** shows a region from the 3D reconstruction of a second specimen, in which the green iso-concentration surfaces encompass regions in the point cloud containing over 6 at. % of C+N, helping to visualize an individual collagen fibril. The composition profile through the collagen fibril calculated along the red cylinder and plotted in **Figure 9B**, allows for estimating a fibril diameter of approx. 3.6 nm that agrees with the (S)TEM observation. The trace elements Mg and Na are enriched at the interface between the collagen fibril and HAP crystals and depleted within the collagen fibrils, **Figure 9C**. Na also shows a significantly higher abundance in the HAP crystal, note there is peak overlap at 23 Da, however, natural abundance of $^{46}Ca^{2+}$ is only 0.004% and therefore the signal can be clearly assigned to Na. The composition profile plotted for the ionic species, **Figure 9D**, shows that the organic signals consist mainly of signals located at 28, 29 and 44 Da, and a mass spectrum extracted only from the inner part of the collagen fibrils shows that the major signals are 12, 28, 44 and 56 Da, corresponding respectively to $C^+$ (12 Da); $CO^+$, $N_2^+$, $CH_2N^+$ (28 Da) or $C_2H_4^+$; $CO_2^+$, $CH_2NO^+$, $CH_4O^+$, $CH_4N_2^+$, $C_2H_6N^+$, $C_3H_8^+$ (44 Da); 56 Da to $CN_2O$+, $N_4^+$, $C_4H_8^+$, $C_3H_4O^+$, $C_2H_4N_2^+$, $C_3H_6N^+$. The organic origin of these signals can be ascertained, an absolute peak interpretation cannot yet be achieved, and the actual atomic composition profiles will strongly depend on the peak assignment.

# 4. Discussion

Biominerals, such as bone or teeth, consist mainly of HAP, where Ca is the main component, in addition to other inorganic elements (P, Mg, Na, ...), along with organic component consisting mainly of type I collagen [79]. However, these composite structures represent a major challenge for analysis by APT, both in terms of data yield and spatial and chemical sensitivity due to their different evaporation field strength.

APT analyses are normally performed at low temperatures, typically 20–60 K, and a high applied electric field to trigger field evaporation, both have a critical effect on sample yield. The heating of the specimen's tip by the laser pulse followed by the cooling through heat conduction along the specimen's main axis results in a thermal pulse. Because of the low thermal conductivity of 0.26 – 0.53 W/m·K and low thermal diffusivity of 0.103 – 0.135 $mm^2/s$ of dried bone samples [58–60], the tail of these thermal pulses extends over long times, which reduce the mass resolution [45]. However, APT analysis of heterogeneous composite material at low temperature can lead to a brittle behavior and premature specimen failure [29]. In addition, high electric fields lead to mechanical stresses that cause premature fracture of APT specimens [55,56]; for bone it has been found that specimens typically fracture at an applied DC voltage of 2.5 – 4 kV [36,37], however, the applied electric field strongly depends on the initial tip radius. In addition, biominerals have poor laser absorption properties, so a higher laser pulse energy is required to achieve field evaporation under electrostatic field conditions, leading to a significant increase in thermal tailing of the signals. The thermal tailing of the $^{40}Ca^{2+}$ signal leads to a loss of information, especially for impurities and trace elements that are indistinguishable below the thermal tailing, which has already been postulated by Langelier et al. for bone samples [36]. To establish the APT technique for biomaterials, these limitations need to be overcome to ensure higher chemical sensitivity and data yield.

## 4.1. Parameters optimization for uncoated specimens

Overall, a higher electrostatic field conditions will result in better compositional accuracy of the Ca/P ratio (**Figure S2**). A higher applied electric field can be achieved either by a higher detection rate or by a lower laser energy, both parameters being in contradiction with the sample yield. However, to increase the sample yield, lower electric fields are favorable, thus higher base temperatures and laser pulse energies are typically favored, both leading to a reduction in the amplitude of the electric field necessary to trigger field evaporation at the required rate. Both results in a more pronounced thermal tailing of the signals and a higher background, which can be reduced by using a lower laser frequency. This will increase the time between laser pulses, providing a longer cooling time, but will also result in a slower data acquisition. However, in general a compromise must be found between the different measurement's parameters, which imposes a trade-off with the spatial and chemical resolution.

The preparation of specimens and their geometry also play an important role. Firstly, a small tip radius (r<60 nm) results in a higher electrostatic field at a given DC voltage compared to a larger tip radius (d>100 nm). However, a smaller tip radius at a constant laser energy leads to a higher peak temperature and potential heat accumulation in a poor thermal conductor [80]. Secondly, a higher sample yield was reported when the orientation of the collagen fibrils is parallel rather than perpendicular to the specimen's main axis [36]. The stress concentration at the interface of the high and low field phases can lead to delamination between the two phases, which has been reported for multilayer systems [47,81]. Thirdly, the sample length (4 µm vs. 2 µm) appears to have a significant effect on the mass spectra for bone measurements (**Figure S6**). Shorter specimens dissipate the thermal energy from the sample material to the more conductive Si support post faster than longer specimens due to the shorter distance, which can significantly reduce background and thermal tailing. In addition, the cryo-preparation has not shown any effect on the data quality in general (**Figure 3**).

For the data quality and accuracy the Ca/P ratio was used, since it's an indicator for the HAP phase which is the major component in bone structures [82]. However, it's debatable whether the Ca/P ratio is a suitable indicator, as a Ca/P deficit between 1.2 – 1.5 is always present in bone samples and scatter between species, age, etc. [83,84]. In addition, the Ca/P ratio has been shown to be strongly dependent on the APT measurement parameters [13,36,44]. It is common for poorly conducting materials to field evaporate as molecular ions rather than atoms; these larger molecules may undergo a process of dissociation into different ions on their way to the detector [52,85]. It is known that this can lead to a loss of neutral species [52], which can strongly influence the composition of oxides, nitrides, etc. [52,85–87]. These dissociation reactions can normally be traced by using a correlation histogram [52]. **Figure S7** shows many dissociation tracks of larger phosphate molecules into smaller fragments, yet no neutral dissociation tracks could be observed. However, it is possible that neutral species do not reach the detector if they are formed shortly after evaporation from the tip surface [85,87].

In general, the high background and thermal tailing of the $^{40}Ca^{2+}$ signal cannot be completely suppressed by optimizing the measurement parameters and/or specimen geometries for bone specimens, resulting always in a loss of information. This makes it impossible to identify and quantify trace elements and complicates the determination of isotope ratios, which can be crucial for the quantification for e.g. for the dating of fossils and minerals [88].

## 4.2. Performance enhancement from metallic coating

To overcome the abovementioned limitations, to improve the chemical sensitivity by increasing the mass resolution and decrease the thermal tailing and background level, it has been reported that the deposition of a thin metallic coating on APT specimens with poor thermal conductivity can be beneficial [46,89,90]. Previous studies have shown that a coated layer can be used to increase the electrical conductivity of the sample, allowing measurements of insulators even in voltage pulse mode [91]. *Ex-situ* coating using sputtering, physical- (PVD) or chemical vapor deposition (CVD) or atomic-layer deposition (ALD) had previously been reported [46,92–94]. Comparison of previous reports of *ex-situ* [95] and *in-situ* [96] coatings on the same material has not shown any advantage of using an *ex-situ* deposition technique. The major drawbacks of *ex-situ* deposition are the need for a second device, in addition the already sharpened specimens are exposed to the ambient environment during transport which can lead to oxidation and may compositional changes in the specimens, and typically there is no way to monitor the coating process with the precision offered by scanning-electron imaging during the *in-situ* coating process. Wang et al. reported an attempt to coat bone specimens with Ag, but no results were reported [97]. Schwarz et al. demonstrated a new method for *in-situ* coating of APT specimens in the FIB chamber, where a homogeneous Cr layer was deposited on pre-sharpened APT specimens [48]. This method resulted in a number of improvements for both conductive and non-conductive materials, such as higher sample yield, mechanical stabilization, reduction of thermal tailing and background, along with evidence of an improvement in mass resolution for hard biomaterials [48].

The improved electrical conductivity of the samples can already be observed in the electron images in **Figure 4**. In APT, the better thermal and electrical conductivity of the bone specimen overall due to the Cr coating results in improved mass resolution, **Figure 5**, a reduction of the background by a factor of 1–2, and in improved capacity for isotopic measurements (**Table 2**). The coating also resulted in a more accurate Ca/P ratio of 1.58, which is closer to the theoretical value of 1.67 (**Table 1**). The reduced thermal tailing and background allows more signals to be detected, making it possible to detect even complete amino acids such as hydroxyproline at 131 Da and glutamic acid at 147 Da at higher masses, which are building blocks of collagen structure, **Figure 5**. However, typical amino acids such as glycine ($C_2H_5NO_2$ at 75 Da), which are 1/3 of the total weight in collagen, do not appear as an obvious signal in the mass spectrum. Whereas the signals of glutamic acid and hydroxyproline, both found in bone collagen up to 10-12% by weight [73], are noticeable. These amino acids could be detected may be due to differences in the fragmentation of different organic molecules during the field evaporation or to differences in the binding and/or stabilization of these molecules. The field evaporation and fragmentation mechanisms for organic molecules in a high electric field are still not fully understood, and these aspects needs to be further explored in the future to ensure more accurate and reliable identification of signals originating from the organic component in biomaterials such as bone. Initial theoretical calculations of the fragmentation of organic molecules during the field evaporation process have been proposed, but these are limited to simple organic molecules, in most cases attached to metallic surfaces as a self-assembled monolayer (SAM) [98–102]. However, these calculations do not include complex structures such as collagen arrangements, inter/intramolecular interactions, etc. and need to be further established in the future to understand the fragmentation of complex organic molecules in a high electric field.

The heterogenous composite bone structure often results in an inhomogeneous and fluctuating detection rate with many "bursts" due to the different evaporation field strengths of HAP and the organic structure. An evaporation field strength of 11 V/nm was determined for HAP [13] and an evaporation field strength of 15 V/nm for bone [36]. For organic structures, little information is available on the evaporation field strengths assumed to be associated with inter- and intramolecular interactions in organic structures. However, organic materials are thought to have a very low evaporation field strength of 3–10 V/nm [99,100]. In both measurements the target detection rate was set to 0.5% (0.005 ions per 100 pulses). Without coating the detection rate varies by ± 0.242%. After Cr coating, the detection rate varies by only ± 0.091%. The metal coating not only improved the mechanical stability and sample yield of the bone specimens, but also achieved a more homogeneous field evaporation, resulting in improved spatial resolution as the depth dimension is build based on the detection sequence.

It has been reported that the metallic coating can influence the effective electric field [91,103] and may also affect the absorption of laser energy. In the future, different coating materials and thicknesses will be further investigated and their effect on the data quality and accuracy of APT measurements of biomaterials, especially bone.

### 4.3. What are the trace elements?

The interpretation of the some of the mass peaks with a low signal-to-background ration is often not straightforward. There are no standards in the literature and the interpretation of the same signals vary across several studies. However, the ability to detect and quantify low-abundance isotopes is essential for the unambiguous detection of trace elements in biominerals in general, e.g. for the investigation of bone diseases (osteolysis, osteo imperfecta), to study the influence of nutrition [104] or external influences (implants, etc.) on the bone structure [38], and plays an important role in tissue engineering and bone diet. APT data was collected from pure HAP and collagen (uncoated), **Figure 6**, to aid in the interpretation and identification of signals in the mass spectrum.

Bone HAP can be substituted by a significant amount of trace elements due to the high number of vacancies, which can accumulate in the bone and alter the solubility and crystal shape, thus affecting the collagen-HAP interaction and directly influencing the physiological properties. Typical substituent cations for $Ca^{2+}$ are $Na^{+}$, $Mg^{2+}$, $Sr^{2+}$, $Al^{3+}$, for $OH^{-}$ are typically anions such as $Cl^{-}$ and $F^{-}$, and carbonate ions $CO_3^{-}$ or citrate ($C_6H_5O_7^{-}$) can replace the hydroxyl ($OH^{-}$) or phosphate ions ($PO_4^{3-}$) [105–110]. APT is in principle ideally suited to detect and quantify trace elements and their local spatial distribution, bringing insights into processes and bonding between organic-inorganic interfaces.

A total of five different trace elements were identified in all 22 samples measured (**Table 3**). Na (1.04 ± 0.47 at. %) and Mg (0.53 ± 0.10 at. %) were detected in all samples. Mg is one of the most important cofactors for the enzymes necessary for the synthesis of the bone matrix [104,111] and Na plays a key role in ion exchange transport during bone formation [104]. In 82% of the samples, K (0.02 ± 0.01 at. %) was detected in very small amounts. K has a beneficial effect, reducing the loss of Ca from the bone matrix and controlling bone density [112]. Mn (0.07 ± 0.02 at. %) was detected in low concentrations in 45% of the samples. Mn has a positive effect on bone and collagen formation [113]. Al was detected in 41% of the samples, with Al ions having a negative effect, interfering with mineralization and the growth and activity of bone cells [114]. The most common substitution ion in bone HAP is carbonate ($CO_3^-$), with approximately 4–8 wt. % of HAP being carbonated [115]. A higher carbon content was observed in the HAP matrix of approximately 11.58 ± 3.37 at. %. APT has been demonstrated to be able to detect even small amounts of trace elements such as K and Mn, which may in the future provide important insights into e.g. aluminum related bone diseases and correlations between the presence of trace elements and basic bone functions and mechanics.

**Table 3**: Analysis of the proportion of trace elements detected by APT in mouse femur specimens. Asterisks indicate the number of samples in which a signal was observed out of 22 measurements. The errors were calculated with the standard deviation.

| | **$Na^+$ (23 Da)** | **$Mg^{2+}$ (12, 12.5 & 13 Da)** | **$Al^{2+}$ (13.5 Da)** | **$K^{2+}$ (19.5 Da)** | **$Mn^{2+}$ (27.5 Da)** |
|---|---|---|---|---|---|
| **concentration (at. %)** | 1.04 ± 0.47 | 0.53 ± 0.010 | 0.03 ± 0.01 | 0.02 ± 0.01 | 0.07 ± 0.02 |
| **Abundance * (%)** | 100 | 100 | 40.91 | 81.82 | 45.46 |

## 4.4. Collagen-Mineral arrangement

Bone has a hierarchical structure across different length scales, from the macro to the nano- and near-atomic scale. Understanding the structure across all length scales is necessary to better understand the complex mineralization mechanisms of this challenging structure and its properties, particularly in relation to disease, bone repair and integration. The smallest major components of this hierarchical structure are HAP crystals and collagen molecules, which consist mainly of the amino acids (hydroxyproline and glycine) [116]. Individual collagen molecules are 300 nm long and form a triple helix structure (tropocollagen) with a diameter of 1.5 nm, which assembles into microfibril bundles of five collagen molecules aligned parallel to a rod-like structure with a diameter of about 4 nm [117]. These microfibrils leave a gap of about 36 nm between the collagen fibrils of the individual collagen stacks in which HAP crystals form. These structures form the building blocks for higher structures of about 0.5 µm, known as collagen fibrils [117,118]. At the microstructural length scale, these fibrils form osteons, which have a lamellar structure, and the Haversian canal, which defines the microstructure [17]. The macrostructure can be divided into an outer cortical (dense) bone structure and an inner trabecular (spongy) bone structure [17]. Several studies [2,15,17–21] have investigated the ultrastructure of bone at different length scales, but chemically-and spatially-resolved imaging at the nanoscale remains challenging and lacking.

The exact mechanism of collagen fibril mineralization has been debated for decades [19–21]. In the early stages of bone growth, HAP crystals are mineralized in the gap zones (interfibrillar mineralization) and grow outwards over the time in the intrafibrillar space [119]. However, the volume between the collagen fibrils is not large enough to explain the volume fraction of HAP in a fully calcified bone, leading to the theory that HAP crystals in the outer region between the collagen fibrils (extrafibrillar mineralization) are mineralized to plate-like crystals along the collagen fibrils, which could explain a large proportion of HAP in the bone structure [19,20].

Our APT analyses of a mouse femur using with the metallic coating showed that we were able to visualize and analyze individual collagen fibrils (**Figure 8** & **Figure 9**). The diameter of the mineralized collagen fibrils was determined to be 3.67 ± 0.67 nm by STEM and 3.61 nm by APT, in good agreement with the literature value of approximately 4 nm [117]. Analysis of the compositional profile through one of the collagens fibrils shows that the interior of the fibril is enriched in organic components and depleted in Ca and P, but there is still a significant residual concentration of Ca and P within the fibril (**Figure 9B-D**). In the extracted mass spectrum within the collagen fibrils (**Figure 9E**), significant signals of Ca and P can be observed in addition to the organic signals consisting mainly of 28, 44 and 46 Da. The signal at 28 Da plays a crucial role in the identification of fibrils, as has been observed in other studies [94] and is predominated compared to signals from carbon. However, the origin of the signal is not yet clear, as it has been identified as CO in all studies [36,37,94], but the following molecules have the same mass-to-charge state ratio $CO^+$, $N_2^+$, $CH_2N^+$ or $C_2H_4^+$, which makes interpretation and compositional analysis difficult. Also the interpretation of small signals is often not clear and straightforward. There are no standards in the literature and the interpretation of the same peak may be different across several studies. Which can lead to differences in the composition analysis. The presence of Ca and P inside the collagen fibrils suggests interfibrillar mineralization. However, organic molecules typically have a lower evaporation field strength [99,100] than inorganic materials such as (HAP), hence enhanced evaporation behavior compared to the inorganic counterpart that lead to local differences in topography and hence electric fields, which can lead to spatial distortions arising from so-called local magnification effects [120,121]. This can lead to the detection of Ca signals within the collagen fibril structure due to the overlap of ion trajectories, causing difficulties in an accurate quantification of interfibrillar mineralization. However, the iso-concentration surfaces in **Figure 8** show that the majority of Ca and P are located between the collagen fibrils, and the compositional profile shows that Ca is the major constituent outside the collagen fibrils (**Figure 9B,D**), suggesting extra-fibrillar mineralization. The fact that the collagen fibrils are not perfectly parallel to each other (**Figure 8**) also suggests some distortion of the fibrils, which could result from extra-fibrillar mineralization as postulated [19,21].

A significant residual C fraction can be observed outside the collagen fibrils, in the HAP crystals, which can be attributed to the fact that biogenic HAP is typically partial carbonated in the bone up to 4–8 wt. % [115]. The compositional profile (**Figure 9C**) shows that Mg is partitioning into the HAP region, indicating substitution in the crystal. Na also shows a predominant accumulation at the interface and a strong increase in concentration in the HAP, contrary to Langelier's [36] observation that Na co-localizes within the microfibrils. The presence of both trace elements at the interface raises the question of whether they originate from specific binding proteins or proteoglycans responsible for adhesion between the collagen fibrils and the HAP crystal, which requires further analysis. Based on our APT results, we conclude, in good agreement with the observation by Lee et al. in human bone [37], that we have found evidence for both mineralization mechanisms present in the mineralization of collagen fibrils, consistent with the recent model by Reznikov et al. [2].

## 5. Conclusion

We have demonstrated a new method using an *in-situ* approach for a metallic coating of bone specimens that allows to increase the sample yield, reduce the mass resolution and background level, resulting in more accurate measurements and composition and overcoming the reported difficulties in general measuring biominerals in APT. The interpretation of small signals is often not clear and straightforward for biominerals, especially with a high organic fraction, whereby comparing the mass spectra of HAP, collagen and bone, many of these signals could be clearly attributed to organic origin. Therefore, individual mineralized collagen fibrils with diameters of 3.67 ± 0.67 nm ((S)TEM) and 3.61 nm (APT) could be visualized and analyzed. A combination of intrafibrillar and extrafibrillar mineralization of collagen fibrils was observed in the mouse femoral structure of these samples. The presence of Ca signals within the collagen fibrils indicates intrafibrillar mineralization, whereas the presence of Ca-rich areas outside the collagen fibrils and the slight distortion of the orientation of the fibrils relative to each other indicate extrafibrillar mineralization. The improved chemical sensitivity allows the detection of several trace elements such as Na, Mg, Al, Mn and K in the bone structure. Na and Mg are enriched at the interface between collagen fibrils and HAP crystals, suggesting the presence of specific binding proteins at the collagen-HAP interface.

Furthermore, the ability to unambiguously detect trace elements in biominerals is generally essential and opens new possibilities for understanding the hierarchical arrangement and chemical heterogeneity of bone structures at the atomic level, providing new, unexplored insights into biomineralization processes and establishing APT in biomaterials research in the future. A better understanding of these biomineralization processes is essential for a better fundamental understanding of the formation of bone structure, the development of bone diseases and bone repair mechanisms.

## Data availability

All data needed to evaluate the conclusions in the paper are included in the paper and/or supplementary materials. Additional data for this work can be requested from the contributed authors.

## Declaration of Competing Interest

The authors declare that they have no known competing financial interest or personal relationships that could have appeared to influence the work reported in this paper.

## Acknowledgments

T.M.S. gratefully acknowledges the financial support of the Walter Benjamin Program of the German Research Foundation (DFG) (Project No. 551061178). T.M. S. M.D. and B.G. are grateful for funding from the DFG through the award of the Leibniz Prize 2020 from B.G.. M.D would like to acknowledge the GPM (Rouen, France). C.J. acknowledges the National Research Foundation of Korea (NRF) grant funded by the Korea government (MSIT) (RS-2024-00359650). A.P. acknowledge funding from the EPSRC funded CDT in Advanced Characterisation of Materials. V.G.G acknowledges the financial support of the EPSRC and SFI Centre for Doctoral Training in the Advanced Characterization of Materials (CDT-ACM) and the La Caixa Foundation Postgraduate Students Fellowship.

Uwe Tezins, Andreas Sturm and Christian Broß are acknowledged for their support to the FIB & APT facilities at MPIE.

We would like to acknowledge Edgar Ibarguen and Tess Boreham at the Central Biomedical Services (CBS) at Imperial College London for their support and generous donation of mice tissue, sourced from residuals of animals previously used in non-skeletal research.

## Author Contributions

**Tim M. Schwarz**: Conceptualization, Data curation, Formel analysis, Investigation, Methodology, Project administration, Validation, Visualization, Funding acquisition Writing – original draft

**Maïtena Dumont**: Validation, Writing – review & editing

**Victoria Garcia-Giner**: Investigation, Writing – review & editing

**Chanwon Jung**: Investigation, Formel analysis, Writing – review & editing

**Alexandra Porter**: Funding acquisition, Resources, Supervision, Writing – review & editing

**Baptiste Gault**: Funding acquisition, Resources, Supervision, Writing – review & editing

## Supplementary Information

Supplementary information is available for this paper.

# Supplementary information

## Advancing atom probe tomography capabilities to understand bone microstructures at near-atomic scale

Tim M. Schwarz[1*], Maïtena Dumont[1,2], Victoria Garcia-Giner[3,4], Chanwon Jung[1,5] Alexandra Porter[3], Baptiste Gault[1,3*]

1. Max-Planck-Institute for Sustainable Materials, Max-Planck-Str. 1, Düsseldorf 40237, Germany
2. *now at* Groupe Physique des Matériaux, Université de Rouen, Saint Etienne du Rouvray, Normandie 76800, France
3. Department of Materials, Imperial College London, London, SW7 2AZ, UK
4. *now at* Rosalind Franklin Institute, Harwell Campus, Didcot, Oxfordshire, OX11 0QX
5. *now at* Department of Materials Science and Engineering, Pukyong National University, 45 Yongso-ro, Nam-gu, 48513 Busan, Republic of Korea

*Corresponding author E-mail address: tim.schwarz@mpie.de, b.gault@mpie.de

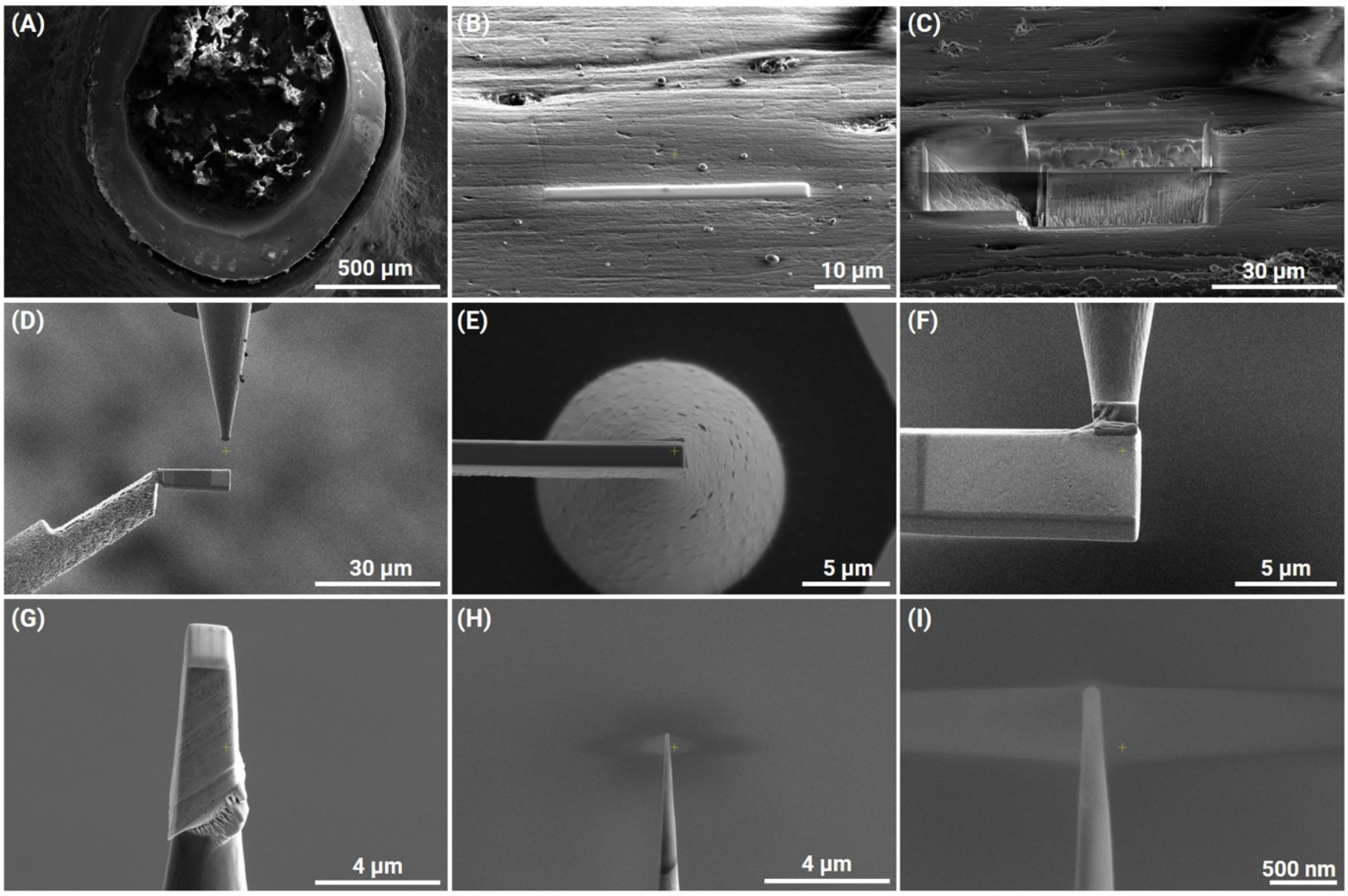

**Figure S 1**: Detailed description of the FIB preparation method used to prepare the samples of the mouse femur specimen. (**A**) The mouse femur bone. (**B**) Pt was first deposited on the selected area and finally three trenches (**C**) were cut into the sample to extract the lamella. In (**D-F**) the lamella is placed and glued onto the Si support post. (**G**) The final mounted lamella on the Si post before the final milling process. (**H, I**) A ring-shaped pattern with decreasing inner diameter was used to shape the sample into a needle-like sample with a tip radius < 100 nm (**I**). The charging of the sample is due to the poor electronic conductivity of the sample (charge-up effect).

**Table S 1:** Comparison peak identification between the mouse femur specimen with and without Cr-coating. There are several possible combinations of ions/molecules for the same signal, the interpretation used is marked in red. Note that there are many more possible combinations for each signal, especially for the organic molecules, which makes an unambiguous interpretation difficult.

| | **uncoated** | | **coated** | | |
|---|---|---|---|---|---|
| **mass-to-charge state ratio (Da)** | **Ion/molecule type** | **charge state** | **mass-to-charge state ratio (Da)** | **Ion/molecule type** | **charge state** |
| 1 | H | + | 1 | H | + |
| 2 | $H_2$ | + | 2 | $H_2$ | + |
| 3 | $H_3$ | + | 3 | $H_3$ | + |
| 6 | C | 2+ | 6 | C | 2+ |
| 7 | N | 2+ | 7 | N | 2+ |
| 12 | Mg, C | 2+/+ | 12 | Mg, C | 2+/+ |
| 12.5 | Mg | 2+ | 12.5 | Mg | 2+ |
| 13 | Mg, CH | 2+/+ | 13 | Mg, CH | 2+/+ |
| 13.5 | Al | 2+ | 13.5 | Al | 2+ |
| 14 | N, $CH_2$ | + | 14 | N, $CH_2$ | + |
| 15 | NH, $CH_3$ | + | 15 | NH, $CH_3$ | + |
| 15.5 | P | 2+ | 15.5 | P | 2+ |
| 16 | O | + | 16 | O | + |
| 17 | OH | + | 17 | OH | + |
| 18 | $OH_2$ | + | 18 | $OH_2$ | + |
| 19 | $OH_3$, F | + | 19 | $OH_3$, F | + |
| 19.5 | K, $C_3H_3$ | 2+/2+ | 19.5 | K, $C_3H_3$ | 2+/2+ |
| 20 | Ca | 2+ | 20 | Ca | 2+ |
| 21 | Ca, $PO_2$ | 3+ | 21 | Ca, $PO_2$ | 3+ |
| 21.5 | ?/Ca | + | 21.5 | ?/Ca | + |
| 22 | Ca | 2+ | 22 | Ca | 2+ |
| 23 | Na | + | 23 | Na | + |
| 24 | Mg | + | 24 | Mg | + |
| 26 | CN | + | 25 | Cr | 2+ |
| 27 | CNH | + | 26 | Cr/CN | 2+ |
| 27.5 | Mn | 2+ | 26.5 | Cr | 2+ |
| 28 | CO | + | 27 | Cr/CNH | + |
| 29 | COH | + | 27.5 | Mn | 2+ |
| 30 | NO | + | 28 | CO | + |
| 31 | P | + | 29 | COH | + |
| 31.5 | $PO_2$ | 2+ | 30 | NO | + |
| 32 | $O_2$ | + | 31 | P | + |
| 33 | $O_2H$ | + | 31.5 | $PO_2$ | 2+ |
| 34 | $O_2H_2$ | + | 32 | $O_2$ | + |
| 38 | C3H2 | + | 33 | CrO | 2+ |
| 39 | $C_3H_3$ | + | 34 | CrO | 2+ |
| 39.5 | $PO_3$ | + | 34.5 | CrO | 2+ |
| 40 | $C_2O$ | + | 35 | CrO | 2+ |
| 40.5 | $PO_3H_2$ | 2+ | 35.5 | ? | - |
| 41 | COH | + | 36 | $C_3$ | + |
| 42 | CNO | + | 37 | $C_3H$ | + |
| 43 | CHNO | + | 38 | $C_3H_2$ | + |
| 44 | $CO_2$ | + | 39 | $C_3H_3$ | + |
| 44.5 | ? | - | 39.5 | $PO_3$ | 2+ |
| 45 | $COH_2$ | + | 40 | $C_2O$ | + |
| 46 | $COH_3$ | + | 40.5 | $PO_3H_2$ | 2+ |
| 47 | PO | + | 41 | COH | + |
| 48 | POH | + | 42 | CNO | + |
| 50 | $C_4H_2$ | + | 42.5 | ? | - |
| 51 | $C_4H_3$ | + | 43 | CHNO | + |
| 52 | $C_2N_2$ | + | 43.5 | ? | - |
| 53 | $C_2HN_2$ | + | 44 | $CO_2$ | + |
| 54 | $C_2H_2N_2$ | + | 45 | $COH_2$ | + |
| 55 | $P_2O_3$ | 2+ | 46 | $COH_3$ | + |
| 56 | CaO | 2+ | 47 | PO | + |
| 57 | $CH_3NO$ | + | 48 | POH | + |
| 58 | $C_3H_6O$ | + | 48.5 | ? | - |
| 59 | $C_3H_7O$ | + | 50 | Cr | + |
| 60 | $C_3H_8O$ | + | 51 | $C_4H_3$ | + |
| 63 | $PO_2$ | + | 52 | Cr | + |
| 64 | $PO_2H$ | + | 53 | Cr | + |
| 65 | $PO_2H_2$ | + | 54 | Cr | + |
| 66 | $PO_2H_3$ | + | 55 | $P_2O_3$ | 2+ |
| 67 | ? | + | 56 | CaO | 2+ |
| 68 | $C_3O_2$ | + | 56.5 | ? | - |
| 69 | $C_3O_2H$/Ga | + | 57 | $CH_3NO$ | + |
| 70 | $C_3O_2H_2$ | + | 58 | $C_3H_6O$ | + |
| 71 | $P_2O_5$ | 2+ | 59 | $C_3H_7O$ | + |
| 71.5 | $P_2O_5H$ | 2+ | 60 | $Cr_2O$/$C_3H_8O$ | 2+/+ |
| 72 | $CaO_2$ | + | 60.5 | $Cr_2O$ | 2+ |
| 76 | $C_2H_6NO_2$ | + | 61 | $Cr_2O$ | 2+ |
| 77 | ? | - | 62 | ? | - |
| 78 | $P_2O$ | + | 63 | $PO_2$ | + |
| 79 | PO3 | + | 64 | $PO_2H$ | + |
| 80 | $PO_3H$ | + | 65 | $PO_2H_2$ | + |
| 81 | $PO_3H_2$ | + | 66 | $PO_2H_3$ | + |
| 82 | $PO_3H_3$ | + | 67 | ? | - |
| 83 | $CaP_2O_4$ | 2+ | 68 | CrO | + |
| 84 | $C_7$ | + | 69 | $C_3O_2H$/Ga | + |
| 87.5 | ? | - | 70 | $C_3O_2H_2$ | + |
| 89 | ? | | 71 | $P_2O_5$ | 2+ |
| 90 | ? | | 71.5 | $P_2O_5H$ | 2+ |
| 91 | $C_6OH_3$ | + | 72 | $CaO_2$ | + |
| 92 | ? | - | 73 | ? | - |
| 94 | ? | - | 74 | ? | - |
| 95 | $PO_4$ | + | 75 | ? | - |
| 96 | $PO_4H$ | + | 76 | $C_2H_6NO_2$ | + |
| 97 | $PO_4H_2$ | + | 77 | ? | - |
| 98 | $PO_4H_3$ | + | 78 | $P_2O$ | + |
| 102.5 | ? | - | 79 | $PO_3$ | + |
| 103 | $P_3O_7H$ | 2+ | 80 | $PO_3H$ | + |
| 104 | ? | - | 81 | $PO_3H_2$ | + |
| 105 | $C_4H_{11}NO_2$ | + | 82 | $PO_3H_3$ | + |
| 106 | $C_4H_{12}NO_2$ | + | 83 | $CaP_2O_4$ | 2+ |
| 107 | $C_7H_9N$ | + | 84 | $C_7H$ | + |
| 119 | $CaPO_3$ | + | 85 | $C_7H_1$ | + |
| 120 | ? | - | 86 | $C_7H_2$ | + |
| 121 | ? | - | 87 | $C_7H_3$ | + |
| 123 | ? | - | 88 | $C_7H_4$ | + |
| 126 | $P_2O_4$ | + | 89 | $C_7H_5$ | + |
| 142 | $P_2O_5$ | + | 90 | $C_7H_6$ | + |
| 143 | $P_2O_5H$ | + | 91 | $C_7H_7$ | + |
| 158 | $P_2O_6$ | + | 92 | $C_7H_8$ | + |
| 159 | $P_2O_6H$ | + | 93 | $C_7H_9$ | + |
| 160 | $P_2O_6H_2$ | + | 94 | $C_7H_{10}$ | + |
| 161 | $P_2O_6H_3$ | + | 95 | $PO_4$ | + |
| 205 | $P_3O_7$ | + | 96 | $PO_4H$ | + |
| 221 | $P_3O_8$ | + | 97 | $PO_4H_2$ | + |
| 222 | $P_3O_8H$ | + | 98 | $PO_4H_3$ | + |
| 223 | $P_3O_8H_2$ | + | 99 | ? | + |
| 284 | $P_4O_{10}$ | + | 100 | $C_6NH_{14}$ | + |
| 285 | $P_4O_{10}H$ | + | 101 | $C_6NH_{15}$ | + |
| | | | 103 | $P_3O_7H$ | 2+ |
| | | | 104 | ? | - |
| | | | 105 | $C_4NO_2H_{11}$ | + |
| | | | 106 | $C_4NO_2H_{12}$ | + |
| | | | 107 | $C_7H_9N$ | + |
| | | | 109 | ? | - |
| | | | 110 | ? | - |
| | | | 111 | ? | - |
| | | | 112 | ? | - |
| | | | 113 | ? | - |
| | | | 114 | ? | - |
| | | | 115 | ? | - |
| | | | 119 | $CaPO_3$ | + |
| | | | 122 | ? | - |
| | | | 123 | ? | - |
| | | | 126 | $P_2O_4$ | + |

| | | | |
|---|---|---|---|
| | 128 | ? | - |
| | 129 | ? | - |
| | 131 | $C_6NO_3H_{13}$ | + |
| | 132 | ? | - |
| | 134 | ? | - |
| | 142 | $P_2O_5$ | + |
| | 143 | $P_2O_5H$ | + |
| | 145 | ? | - |
| | 146 | ? | + |
| | 147 | $C_6NO_4H_9$ | - |
| | 152 | ? | - |
| | 158 | $P_2O_6$ | + |
| | 159 | $P_2O_6H$ | + |
| | 160 | $P_2O_6H_2$ | + |
| | 161 | $P_2O_6H_3$ | + |
| | 163 | ? | - |
| | 164 | ? | - |
| | 165 | ? | - |
| | 167 | ? | - |
| | 168 | ? | - |
| | 175 | ? | - |
| | 205 | $P_3O_7$ | + |
| | 210 | ? | - |
| | 221 | $P_3O_8$ | + |
| | 222 | $P_3O_8H$ | + |
| | 223 | $P_3O_8H_2$ | + |
| | 227 | ? | - |
| | 261 | ? | - |
| | 284 | $P_4O_{10}$ | + |
| | 285 | $P_4O_{10}H$ | + |

The full-width half-maximum (FWHM) of the $Ca^{2+}$ peak at 20 Da was measured to estimate the mass resolution. The charge-state ratio (CSR) of $PO_3^{2+}/PO_3^{+}$ [1] and $O_2^{2+}$-$O^{+}$ /$O_2^{+}$ [2] were determined as they reflect the effective amplitude of the electrostatic field at the specimen's tip [3–5]. The Ca/P ratio is widely used as an indicator to identify HAP [6]. The fraction of inorganic (C, P, O, trace elements) to organic (C, N) compounds was calculated based on the peak identification given in **Table S1**.

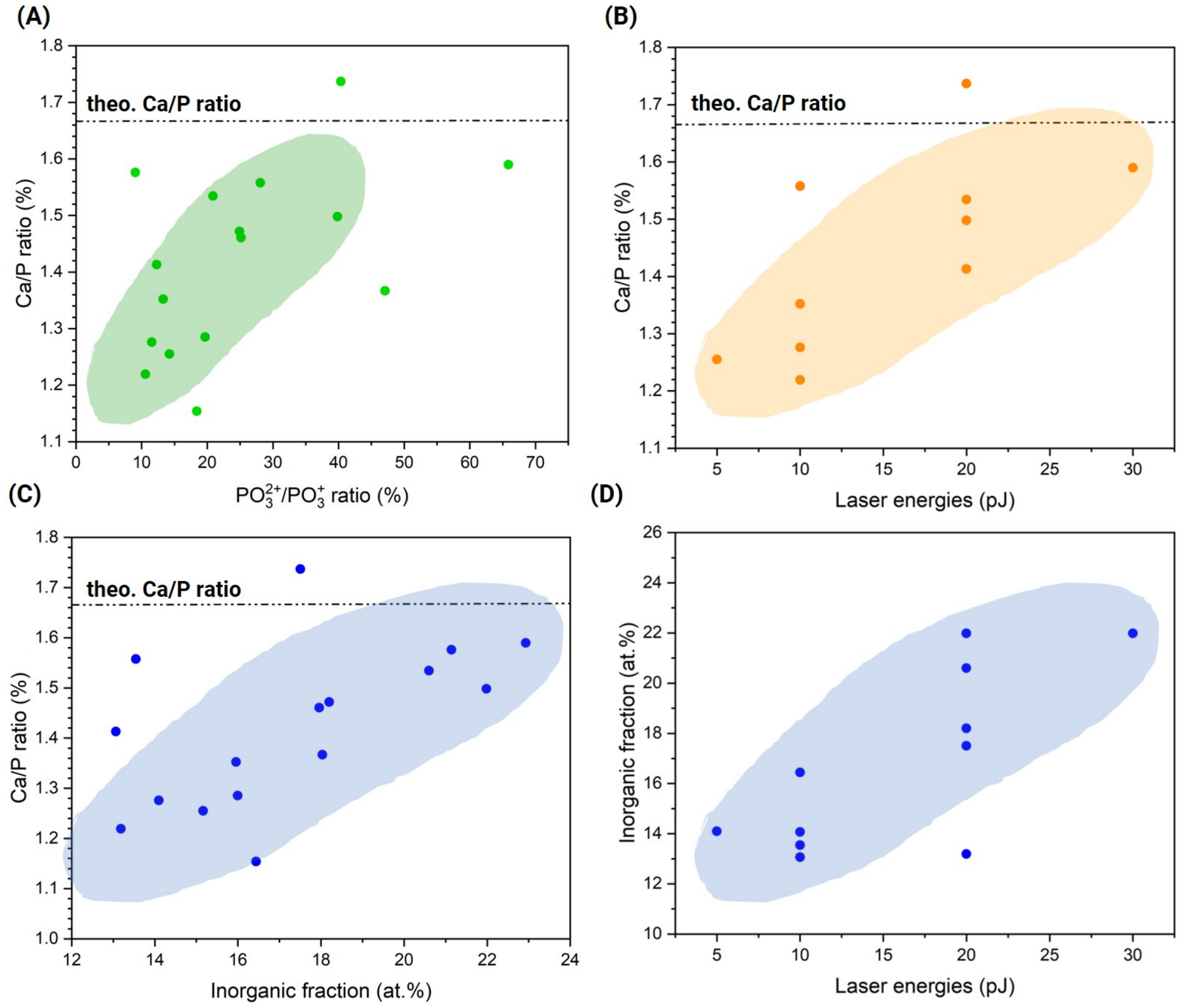

**Figure S 2**: Correlation plot in (**A**) for the Ca/P ratio as a function of $PO_3^{2+}/PO_3^{+}$ CSR, in (B) for the laser energy and in (C) for the inorganic fraction. (D) Correlation between the inorganic fraction and laser pulse energy.

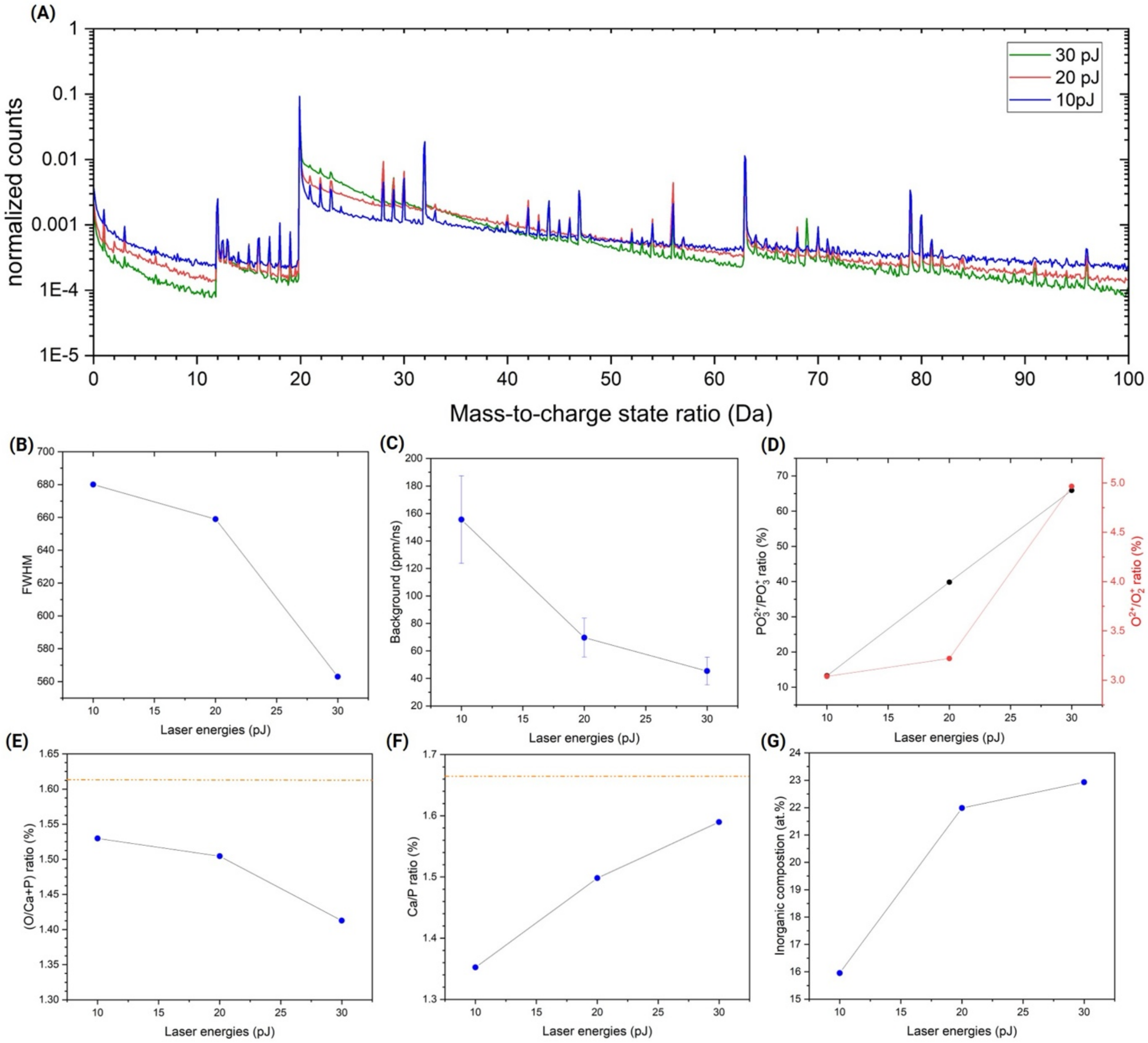

**Figure S 3**: (**A**) Mass spectra of the uncoated mouse femur specimen with different laser energies from 10 pJ to 30 pJ at constant detection rate and laser pulse frequency. (**B-G**) Correlation plots between mass resolution (FWHM), background, Ca/P ratio, O/(Ca+P) ratio, CSR and organic fraction.

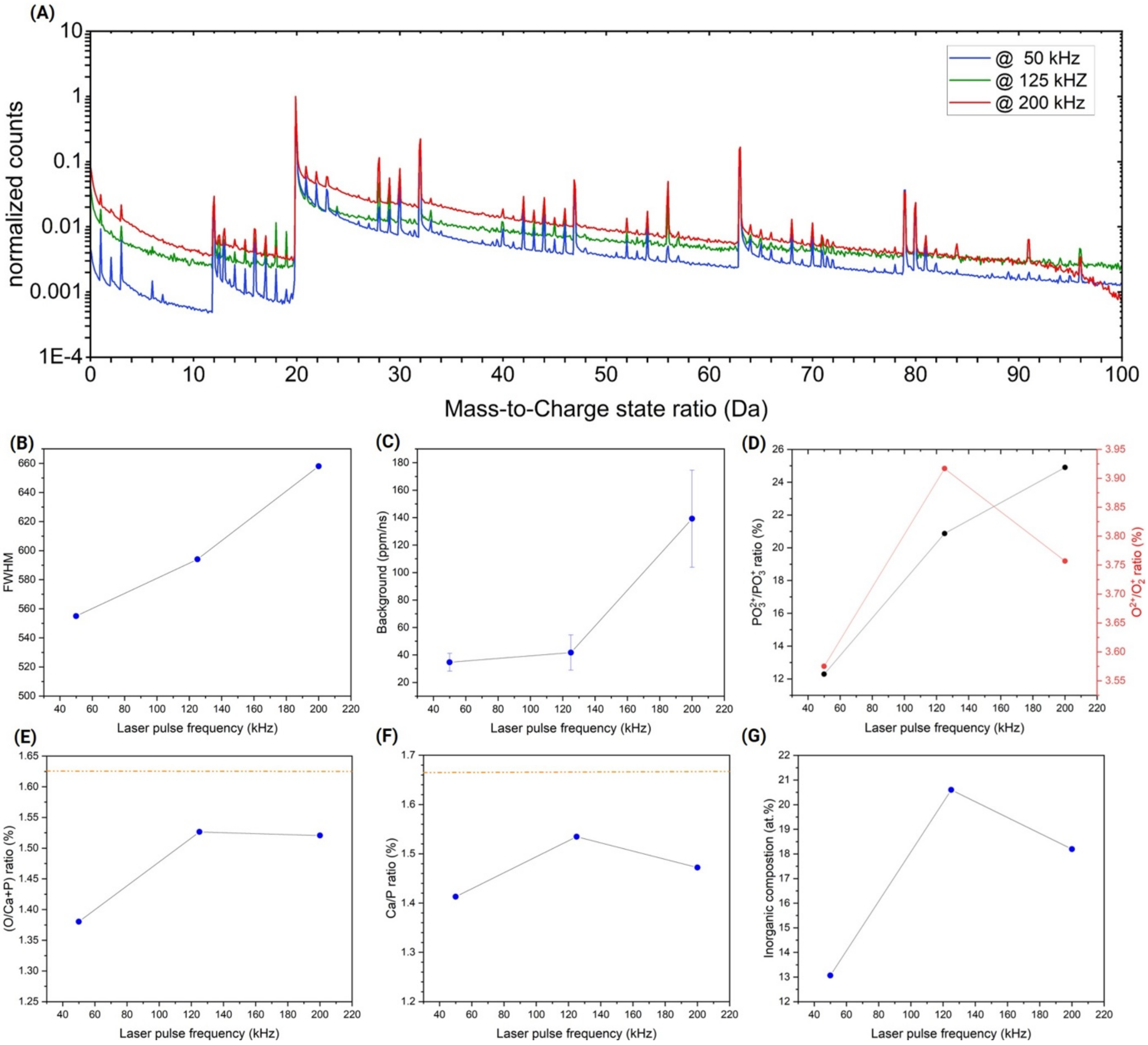

**Figure S 4**: (**A**) Mass spectra of the uncoated mouse femur specimen with different laser pulse frequencies at constant detection rate and laser pulse energy. (**B-G**) Correlation plots between mass resolution (FWHM), background, Ca/P ratio, O/(Ca+P) ratio, CSR and organic fraction.

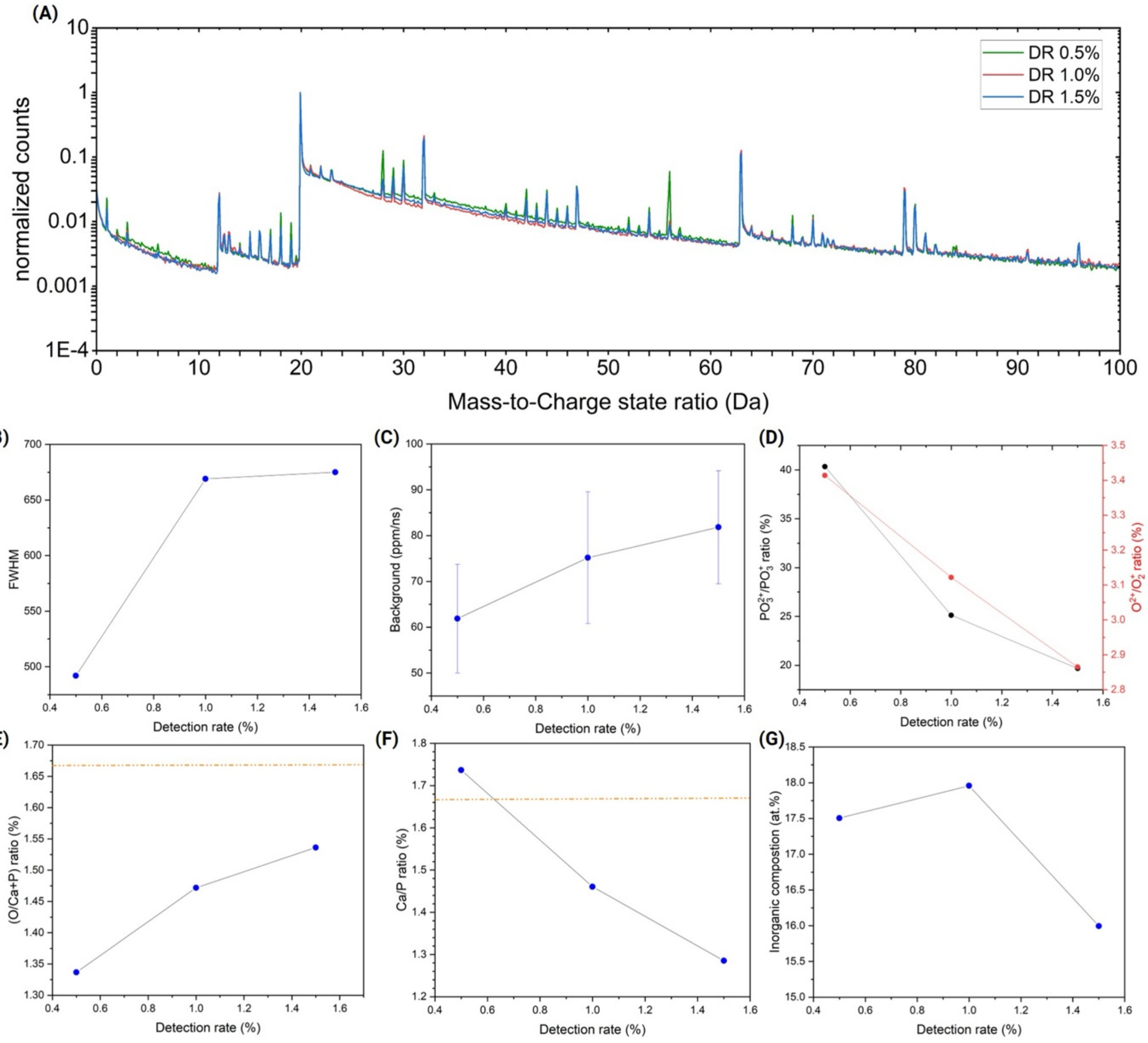

**Figure S 5**: (**A**) Mass spectra of the uncoated mouse femur specimen with different detection rates from 0.5 to 1.5% at constant laser pulse frequency and energy. (**B-G**) Correlation plots between mass resolution (FWHM), background, Ca/P ratio, O/(Ca+P) ratio, CSR and organic fraction.

**Table S 2**: Peak identification in the pure HAP sample. Several carbon-containing signals were observed, indicating some carbon impurities in the synthesized HAP particles.

| mass-to-charge state ratio (Da) | Ion/molecule type | charge state |
|---|---|---|
| 1 | H | + |
| 2 | $H_2$ | + |
| 3 | $H_3$ | + |
| 12 | C | + |
| 15 | NH | + |
| 16 | O | + |
| 17 | OH | + |
| 18 | $OH_2$ | + |
| 19 | $OH_3$ | + |
| 20 | Ca | 2+ |
| 21 | Ca/$PO_2$ | 3+ |
| 22 | Ca | 2+ |
| 27 | CNH | + |
| 28 | CO | + |
| 29 | NO | + |
| 30 | $COH_2$/NO | + |
| 32 | $O_2$ | + |
| 33 | $O_2H$ | + |
| 39 | $P_2O$ | 2+ |
| 40 | Ca | + |
| 41 | $C_2OH$ | + |
| 42 | CNO | + |
| 43 | CNOH | + |
| 44 | $CO_2$ | + |
| 47 | PO | + |
| 55 | $P_2O_3$ | 2+ |
| 56 | CaO | + |
| 57 | CaOH | + |
| 59 | CaF | + |
| 63 | $PO_2$ | + |
| 72 | $CaO_2$ | + |
| 79 | $PO_3$ | + |
| 80 | $PO_3H$ | + |
| 81 | $PO_3H_2$ | + |
| 96 | $PO_4H$ | + |
| 119 | $CaPO_3$ | + |
| 142 | $P_2O_5$ | + |
| 143 | $P_2O_5H$ | + |

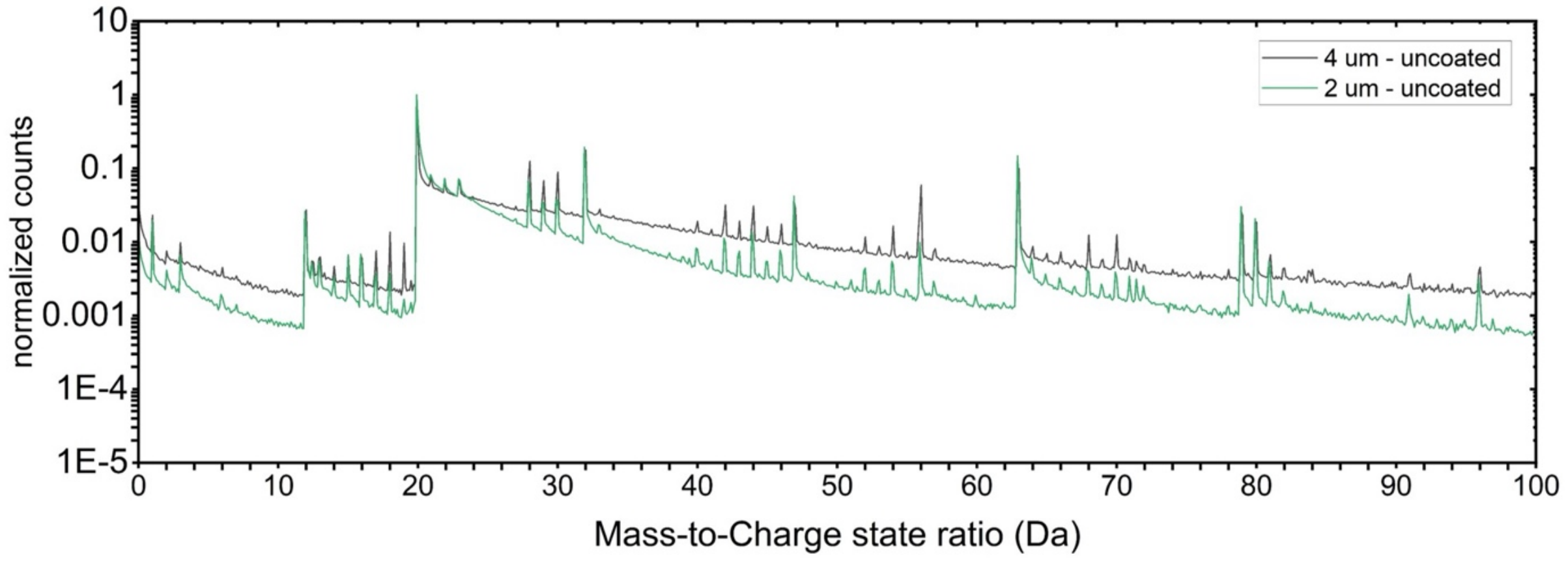

**Figure S 6**: Comparison of the mass spectra of the femur sample with different sample lengths of 4 µm and 2 µm. Both spectra are normalized to the highest peak.

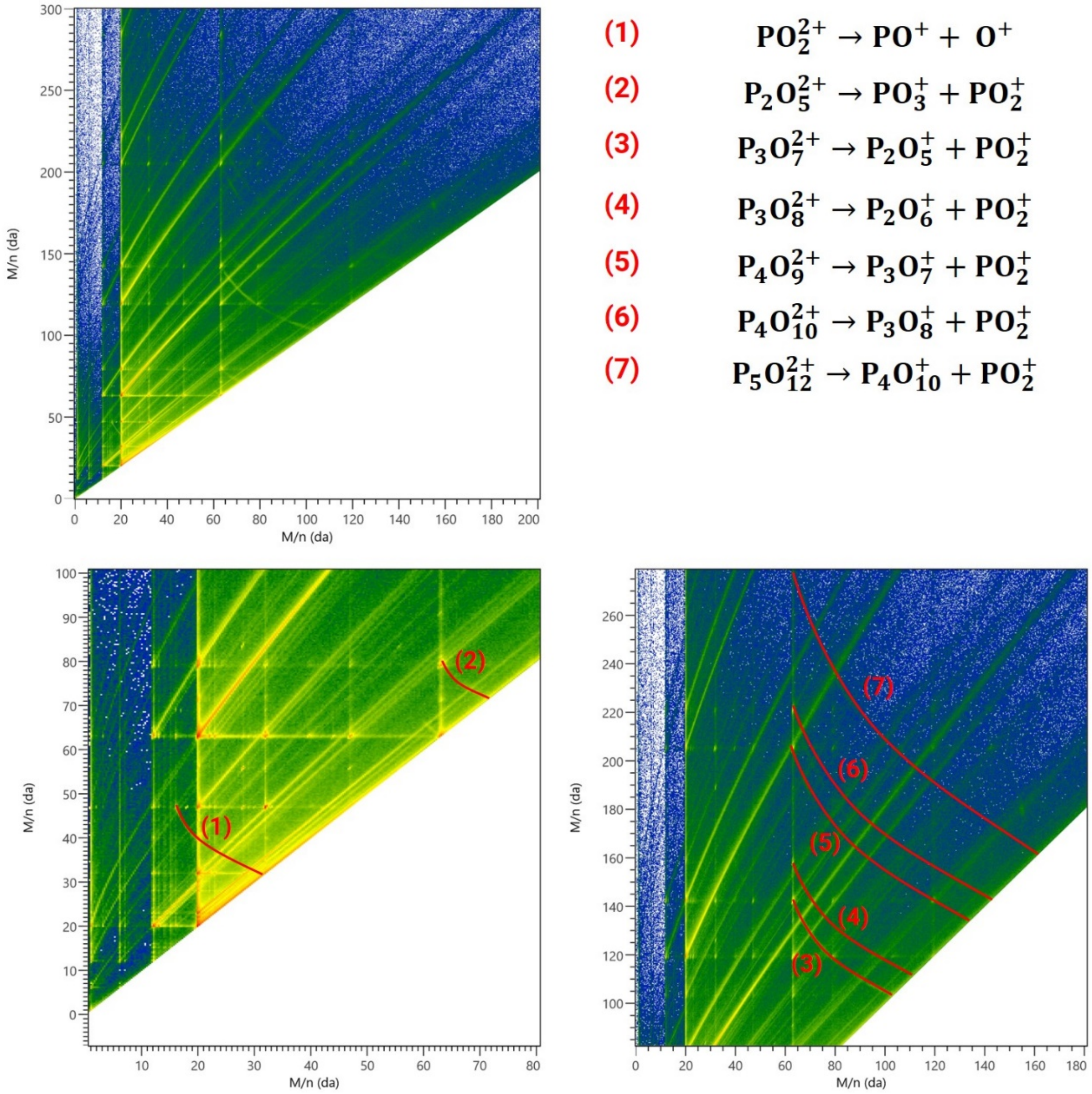

**Figure S 7**: Correlation plot, the dissociation traces are marked with red lines. The vertical and horizontal lines correlate with the field evaporation triggered by the laser pulse. A high intensity of delayed evaporation can be observed, i.e. shortly after the laser pulse hits the specimen atoms/ions continuously field evaporate, which is associated with the poor thermal properties.